\documentclass[aps,twocolumn,showpacs,preprintnumbers,amsmath,amssymb,superscriptaddress,floatfix,nofootinbib]{revtex4}
\usepackage{lipsum}
\usepackage{graphicx}
\usepackage{epsfig}
\usepackage{epstopdf}
\usepackage{multirow}
\usepackage{array}
\usepackage{color}
\usepackage[colorlinks=true,urlcolor=blue,linkcolor=blue,citecolor=blue]{hyperref}
\usepackage{hyperref}
\usepackage{amsmath}
\usepackage{amsfonts}
\usepackage{amssymb}
\usepackage{comment}

\def \nn {\nonumber}
\def \) {\right)}
\def \( {\left(}
\def \] {\right]}
\def \[ {\left[}

\begin{document}


\title{Neutron-proton scattering at next-to-next-to-leading order\\ in Nuclear Lattice Effective Field Theory}

\author{Jose~Manuel~Alarc\'on} 
\affiliation{Helmholtz-Institut f\"ur Strahlen- und
             Kernphysik and Bethe Center for Theoretical Physics,
             Universit\"at Bonn,  D-53115 Bonn, Germany}  
             
\affiliation{Theory Center, Thomas Jefferson National Accelerator Facility, Newport News, VA 23606, USA}

\author{Dechuan~Du}
\affiliation{Institute~for~Advanced~Simulation, Institut~f\"{u}r~Kernphysik,
and J\"{u}lich~Center~for~Hadron~Physics,~Forschungszentrum~J\"{u}lich,
D-52425~J\"{u}lich, Germany}       

\author{Nico~Klein} 
\affiliation{Helmholtz-Institut f\"ur Strahlen- und
             Kernphysik and Bethe Center for Theoretical Physics,
             Universit\"at Bonn,  D-53115 Bonn, Germany}  
             
             \author{Timo~A.~L\"{a}hde}
\affiliation{Institute~for~Advanced~Simulation, Institut~f\"{u}r~Kernphysik,
and J\"{u}lich~Center~for~Hadron~Physics,~Forschungszentrum~J\"{u}lich,
D-52425~J\"{u}lich, Germany}     

\author{Dean~Lee}
\affiliation{Department~of~Physics, North~Carolina~State~University, Raleigh,
NC~27695, USA}

\author{Ning Li}
\affiliation{Institute~for~Advanced~Simulation, Institut~f\"{u}r~Kernphysik,
and J\"{u}lich~Center~for~Hadron~Physics,~Forschungszentrum~J\"{u}lich,
D-52425~J\"{u}lich, Germany}

\author{Bing-Nan~Lu}
\affiliation{Institute~for~Advanced~Simulation, Institut~f\"{u}r~Kernphysik,
and J\"{u}lich~Center~for~Hadron~Physics,~Forschungszentrum~J\"{u}lich,
D-52425~J\"{u}lich, Germany}

\author{Thomas Luu} 
\affiliation{Institute~for~Advanced~Simulation, Institut~f\"{u}r~Kernphysik,
and J\"{u}lich~Center~for~Hadron~Physics,~Forschungszentrum~J\"{u}lich,
D-52425~J\"{u}lich, Germany}   
                 
\author{Ulf-G.~Mei{\ss}ner}  
\affiliation{Helmholtz-Institut f\"ur Strahlen- und
             Kernphysik and Bethe Center for Theoretical Physics,
             Universit\"at Bonn,  D-53115 Bonn, Germany}  
\affiliation{Institute~for~Advanced~Simulation, Institut~f\"{u}r~Kernphysik,
and J\"{u}lich~Center~for~Hadron~Physics,~Forschungszentrum~J\"{u}lich,
D-52425~J\"{u}lich, Germany}     
\affiliation{JARA~-~High~Performance~Computing, Forschungszentrum~J\"{u}lich,
D-52425 J\"{u}lich,~Germany} 

           
\date{\today}

\begin{abstract}
We present a systematic study of neutron-proton scattering in Nuclear Lattice Effective Field Theory (NLEFT), in terms of the computationally efficient
radial Hamiltonian method. Our leading-order (LO) interaction consists of smeared, local contact terms
and static one-pion exchange. We show results for a fully non-perturbative analysis up to next-to-next-to-leading order (NNLO), followed by a 
perturbative treatment of contributions beyond LO. The latter analysis anticipates practical Monte Carlo simulations of heavier nuclei.  
We explore how our results depend on the lattice spacing $a$, and estimate sources of uncertainty in the determination of the low-energy constants of
the next-to-leading-order (NLO) two-nucleon force. We give results for lattice spacings ranging from $a = 1.97$~fm down to $a = 0.98$~fm, and discuss
the effects of lattice artifacts on the scattering observables. At $a = 0.98$~fm, lattice artifacts appear small, and our NNLO results agree well with 
the Nijmegen partial-wave analysis for $S$-wave and $P$-wave channels. We expect the peripheral partial waves to be equally well described once
the lattice momenta in the pion-nucleon coupling are taken to coincide with the continuum dispersion relation, and higher-order (N3LO) contributions
are included. We stress that for center-of-mass momenta below 100~MeV, the physics of the two-nucleon system is independent of the lattice
spacing.
\end{abstract}

\maketitle


\section{Introduction \label{introduction}}

Nuclear Lattice Effective Field Theory (NLEFT) has recently gained prominence as an {\it ab initio} method for the study of 
nuclear structure formation  at low energies.
The advent of NLEFT has largely been due to rapid developments in computational algorithms and resources, which have enabled the efficient combination
of lattice Monte Carlo methods with the low-energy effective field theory of QCD, known as Chiral Perturbation Theory or Chiral Effective Field Theory. 
Such progress has greatly  increased our ability to 
exploit the advantages of the EFT method in the realm of many-body nuclear physics, which remains a highly challenging area of study. Hence, 
impressive progress has been made within NLEFT in furthering our understanding of the spectra, structure and scattering of light- and medium-mass 
nuclei~\cite{Epelbaum:2011md,Epelbaum:2012qn,Epelbaum:2012iu,Epelbaum:2013paa,Bour:2014bxa,Elhatisari:2015iga,Elhatisari:2016owd}, 
see also Ref.~\cite{Lee:2008fa} for an early review.

Chiral EFT provides a model-independent approach to hadronic interactions at the energy scales of interest for nuclear physics. Based on 
the spontaneous and explicit chiral symmetry breaking of QCD, Chiral EFT provides a systematic treatment of such interactions in terms of 
a generic soft scale ($Q$) which is commonly taken to refer to the Goldstone boson mass (such as the pion mass $M_\pi^{}$) or to external 
nucleon momenta. In the nuclear physics context, the EFT is used to work out the interaction potential between the nuclear constituents.
These chiral potentials are then used in an appropriate framework to generate the bound and scattering states.   For the case of the 
nucleon-nucleon (NN) interaction considered here, Chiral EFT
also clarifies the observed hierarchy between many-body contributions to the nuclear force. This power counting can be expressed in terms of 
$Q/\Lambda_\chi^{}$, where $\Lambda_\chi^{}$ refers to the hard scale at which chiral symmetry is restored~\cite{Weinberg:1990rz}. 
The contributions to the NN force are then classified as leading order (LO) for $(Q/\Lambda_\chi)^0$, followed by next-to-leading order (NLO) for
$(Q/\Lambda_\chi)^2$, and next-to-next-to-leading order (NNLO) for $(Q/\Lambda_\chi)^3$ {\it etc.}, in decreasing order of importance. For a recent review of
Chiral EFT in nuclear physics, see Ref.~\cite{Epelbaum:2008ga}. It should also be noted that Chiral EFT provides a method to systematically estimate 
the uncertainty of a calculation at a given 
order in the EFT expansion, which is of special relevance is searches of physics beyond the Standard Model (BSM). With the advent of precision experiments
searching for BSM physics, the importance of well-controlled error estimates for the nuclear contributions have become essential for the
statistical interpretation of purported BSM signals and hence, ultimately, for any claim of detection, see e.g.~Ref.~\cite{Hoferichter:2016nvd}
(and references therein).

The fundamental problem of neutron-proton scattering in NLEFT was first studied at LO in Ref.~\cite{Borasoy:2006qn}, and later 
extended to  NLO in Ref.~\cite{Borasoy:2007vi}, with phase shifts and mixing angles calculated on the lattice using the so-called spherical wall 
method~\cite{Borasoy:2007vy}. This method was used earlier in the context of variational calculations of resonant states in $^4$He~\cite{Carlson:1984zz}.
We shall here revisit, in a systematical manner, the calculation of neutron-proton scattering observables, which also serve to determine the 
low-energy constants (LECs) of the NLO contact terms of NLEFT. Our work is based on an improvement of the spherical wall method known as the 
radial Hamiltonian formalism, which was proposed and pioneered in Ref.~\cite{Elhatisari:2015iga} in the study of alpha-alpha scattering on the lattice. 
In this formalism,  the two-nucleon problem is formulated in terms of radial coordinates for each partial wave. Specifically, lattice points
with the same radial coordinate are grouped together and weighted by the appropriate spherical harmonics, which eliminates the
need to work with a computationally costly $L^3 \times L^3$ problem (where $L$ is the linear dimension of the cubic lattice) without
loss of precision. This approach can be further accelerated by binning lattice points with similar radial coordinates into segments
of width $a_R^{}$, as proposed in Ref.~\cite{Elhatisari:2016hby}.
The advantages of the radial Hamiltonian method were already demonstrated in Ref.~\cite{Lu:2015riz} for the phase shifts and
mixing angles of a system of two nucleons with a simplified  model potential.

In the present work, we address the task of determining the LECs of the two-nucleon force at NLO and NNLO in NLEFT, by means 
of a chi-square minimization with respect to neutron-proton phase shifts and mixing angles. This procedure also allows us to provide quantitative 
estimates of the uncertainties of the NLO constants in
NLEFT, along with estimates of their systematical errors and the impact of such errors on the binding energies of nuclei. It should be noted that 
the pioneering calculations
of Refs.~\cite{Borasoy:2006qn,Borasoy:2007vi} (and almost all calcuations of nuclear properties)
were performed with a coarse lattice spacing of $a  = 1.97$~fm, which corresponds to a relatively low 
momentum cutoff of $\pi / a = 314$~MeV~\footnote{Note that such soft nucleon-nucleon interactions lead to better convergence properties in 
the calculations of many-nucleon systems
and nuclear matter, see e.g.~\cite{Bogner:2005sn}.}. 
Here, we now also study the effects of decreasing the lattice spacing to $a \simeq 1$~fm, 
which greatly decreases the impact of lattice artifacts and systematical errors, and discuss the possibility of further improving the 
lattice action to decrease remaining discretization effects. Note that a first study
of discretization errors and lattice spacing variation at LO has been performed in Ref.~\cite{Klein:2015vna}. Finally, our study of 
lattice spacing variation requires that the
two-pion exchange potential (TPEP) is explicitly accounted for. In prior work at $a  = 1.97$~fm, the TPEP at NLO and NNLO 
contributions were integrated out by means of a 
Taylor expansion in powers of $q^2 / (4 M_\pi^{})$. Since we now use lattice spacings as small as $a \simeq 1$~fm, we need to include the 
full structure of the TPEP in our analysis.

Our paper is organized as follows: The lattice EFT formalism, the radial Hamiltonian method, and the NLEFT potentials up to NNLO are 
presented in Section~\ref{formalism}. In Section~\ref{results}, we give the results of a fully non-perturbative calculation of coupled-channel
neutron-proton scattering up to NNLO, followed by a treatment where the NLO and NNLO contributions are computed perturbatively. We also
study the lattice spacing dependence of the calculated phase shifts and mixing angles. Furthermore, we investigate how the uncertainty
in the four-nucleon LECs propagates into the prediction of nuclear ground-state energies.
In Section~\ref{summary}, we conclude with a brief discussion of planned N3LO calculations and other future directions.


\section{Lattice formalism \label{formalism}}

We begin with a detailed description of the NLEFT lattice Hamiltonian on which our calculations are based.
We denote the (spatial) lattice spacing by $a$, the temporal lattice spacing by $a_t^{}$, and we also define
$\alpha_t^{} \equiv a_t^{} / a$. Our lattice is a periodic cube of volume $L^3$. For non-zero temporal lattice spacing,
we define the transfer matrix as~\cite{Borasoy:2006qn} 
\begin{equation}
M \equiv \: : \exp \left(-\alpha_t^{} H \right):,
\label{transfer_matrix}
\end{equation}
with the Hamiltonian
\begin{equation}
H \equiv H_{\mathrm{free}}^{} + V_{\mathrm{LO}}^{} + V_{\mathrm{NLO}}^{} + \ldots,
\end{equation}
where $H_{\mathrm{free}}$ is the free nucleon Hamiltonian
and $V_{\mathrm{LO}}$, $V_{\mathrm{NLO}}$, {\it etc.}\ contain nucleon-nucleon interactions of progressively higher order
in NLEFT. The colons in Eq.~(\ref{transfer_matrix}) denote normal ordering.
The energy eigenvalues are given by
\begin{equation}
E_\lambda^{} = -\frac{1}{a_t^{}} \log \lambda, 
\end{equation}
where $\lambda$ denotes an eigenvalue of $M$.

Following Ref.~\cite{Elhatisari:2015iga}, we construct the transfer 
matrix in radial coordinates. Specifically, we group the lattice points $(n_x, n_y, n_z)$ with the same radial coordinate,  
by weighting them with the spherical harmonics. Thus, instead of working with the full basis $|\vec R \rangle$, 
one obtains the reduced basis
\begin{equation}
|R \rangle \equiv  \sum_{\vec R^\prime} Y_{l, l_z}^{}(\hat{R}^\prime) 
\delta_{R, R^\prime}^{} |\vec{R}^\prime \rangle,
\end{equation}
where $Y_{l, l_z}^{}$ is the spherical harmonic for angular momentum quantum numbers $(l, l_z)$ and $\delta$ 
denotes the Kronecker delta. We thus obtain the radial transfer matrix
\begin{equation}
M_{\vec{R}^\prime, \vec{R}} \to \tilde{M}_{R^\prime, R}^{}\ .
\end{equation}
A similar approach with a refined grid for the radial lattice was performed in~\cite{Elhatisari:2016hby}.

To determine phase shifts and mixing angles, we apply the method proposed in Ref.~\cite{Lu:2015riz}, whereby these are extracted
directly from the radial wave functions. Specifically, one defines three
radii, $R_\mathrm{in}$, $R_\mathrm{out}$ and $R_\mathrm{wall}$. 
The NN interaction contributes in the range $0<r<R_{\mathrm{in}}$,
while an infinite spherical wall barrier is applied for $r > R_{\mathrm{wall}}$. In the range 
$R_{\mathrm{in}}<r <R_{\mathrm{out}}$, the NN interaction vanishes and the wave function 
can be expanded as a linear combination of the spherical Bessel and Neumann functions, according do
\begin{eqnarray}
\psi_l^{}(r) = A j_l^{}(qr) + B n_l^{}(qr),
\end{eqnarray} 
from which phase shifts and mixing angles can be extracted. For more details, see Ref.~\cite{Lu:2015riz} and the earlier work of Ref.~\cite{Borasoy:2007vy}.
Typical values used later are $R_{\rm in} \simeq 24$~fm, $R_{\rm out} \simeq 28$~fm, and $R_{\rm wall} \simeq 30$~fm.

We shall now give a detailed description of the Hamiltonian $H$, and its various contributions.
The free nucleon Hamiltonian is given by~\cite{Borasoy:2006qn}
\begin{eqnarray}
H_{\mathrm{free}}^{} &\equiv& \frac{3\omega_0^{}}{m_N^{}} 
\sum_{\vec n} \sum_{i, j = 0, 1} a_{i, j}^\dag(\vec{n}) a_{i, j}^{} (\vec{n}) \\
&-& \frac{\omega_1^{}}{2m_N^{}} \sum_{\vec{n}} \sum_{l = 1}^3 \sum_{i, j= 0, 1} \nn \\
&& \times \Big[  a_{i, j}^\dag(\vec{n}) a_{i, j}^{} (\vec n + \hat e_l^{}) + a_{i, j}^\dag(\vec{n}) a_{i, j}^{} (\vec n - \hat e_l^{}) \Big] \nn \\
&+& \frac{\omega_2^{}}{2m_N^{}}\sum_{\vec{n}} \sum_{l = 1}^3 \sum_{i, j=0, 1}  \nn \\
&& \times \Big[ a_{i, j}^\dag(\vec{n}) a_{i, j}(\vec{n} + 2 \hat e_l^{}) + a_{i, j}^\dag(\vec{n}) a_{i, j}(\vec{n} - 2 \hat e_l^{})\Big] \nn  \\
&-& \frac{\omega_3^{}}{2m_N^{}} \sum_{\vec{n}} \sum_{l = 1}^3 \sum_{i, j= 0, 1} \nn \\
&& \times \Big[ a_{i, j}^\dag (\vec{n})a_{i, j}(\vec{n} + 3 \hat e_l^{}) + a_{i, j}^\dag (\vec{n})a_{i, j}(\vec{n} - 3 \hat e_l^{}) \Big], \nn
\end{eqnarray}
where the $\hat e_l^{}$ with $l = 1,2,3$ 
are unit vectors in the spatial directions, and $m_N^{}$ is the nucleon mass. 
In Table~\ref{hopping}, we give the hopping coefficients $\omega_k^{}$ for lattice actions up to $\mathcal{O}(a^4)$. 
Throughout our work, we use the so-called stretched action which is defined in terms of the $\mathcal{O}(a^4)$- 
and $\mathcal{O}(a^2)$-improved actions~\cite{Lee:2008fa}. This gives the stretched hopping coefficients
\begin{eqnarray}
\omega_k^{\mathrm{str}} \equiv \omega_k^{\mathcal{O}(a^4)} + \mathcal{N} \left( \omega_k^{\mathcal{O}(a^4)} - \omega_k^{\mathcal{O}(a^2)} \right),
\end{eqnarray}
where $\mathcal{N} =10$ is adopted in the present calculations.  


\begin{table}[t]
\begin{center}
\caption{Hopping coefficients $\omega_i^{}$ for the free nucleon action, for different levels of improvement. 
\label{hopping}}
\smallskip
\begin{tabular*}{8.5cm}{@{\extracolsep{\fill}}cccc}
\hline \hline
\noalign{\smallskip}
                         &                unimproved             &    $\mathcal{O}(a^2)$ improved         &    $\mathcal{O}(a^4)$ improved 
                         \smallskip \\
\hline
$\omega_0^{}$     &                      1                       &         $5/4$                                &     $49/36$                 \\
$\omega_1^{}$     &                      1                       &         $4/3$                                &     $3/2$                     \\
$\omega_2^{}$     &                      0                       &         $1/12$                              &     $3/20$                   \\
$\omega_3^{}$     &                      0                       &               0                                 &     $1/90$                  \\
\hline \hline        
\end{tabular*}
\end{center}
\end{table}


In Chiral EFT, the NN force is decomposed into the long-range components arising from the 
exchange of pions, and short-range contributions described by contact interactions with increasing powers of momenta.
Such two-nucleon contact operators introduce unknown coefficients which we determine by fitting the data on neutron-proton phase shifts
and mixing angles. In what follows, we present our contact and pion exchange operators.
 
\subsection{Contact interactions}

We begin our treatment of the lattice Chiral EFT interaction by considering the various
contact operators that appear up to NNLO in the chiral expansion. At LO, we consider the following operators 
\begin{equation}
\mathcal{O}_1^{(0)} \equiv \frac{1}{2}  :\sum_{\vec{n}} \rho(\vec{n}) \rho(\vec{n}): , 
\label{contact:LO_a}
\end{equation}
and 
\begin{equation}
\mathcal{O}_2^{(0)} \equiv \frac{1}{2}  :\sum_{\vec{n}} \sum_I \rho_I^{}(\vec{n}) \rho_I^{}(\vec{n}): , 
\label{contact:LO_b}
\end{equation}
as the independent contact operators, with coefficients 
$C$ and $C_I^{}$, respectively. 
Here, $\rho(\vec{n})$ and $\rho_I^{}(\vec{n})$ are the local density and local isospin density operators on 
the lattice, which are defined in App.~\ref{app_operators}. At LO, the coefficients $C$ and $C_I^{}$ are determined
by the spin-singlet ($^1S_0$) and the spin-triplet ($^3S_1$) $S$-wave channels, and can be parameterized as
\begin{equation}
\Bigg[
\begin{array}{c}
C \vspace{.1cm} \\ C_I^{} \\
\end{array}
\Bigg]
= \frac{1}{4}
\Bigg[
\begin{array}{rr}
3  & 1 \vspace{.1cm} \\ 
1 & -1 \\
\end{array}
\Bigg]
\Bigg[
\begin{array}{c}
C_{^1S_0^{}}^{} \vspace{.1cm} \\ C_{^3S_1^{}}^{} \\
\end{array}
\Bigg],
\end{equation}
where $C_{^1S_0^{}}$ and $C_{^3S_1^{}}$ are determined by fitting scattering data in the $^1S_0$ and $^3S_1$ channels.

In Ref.~\cite{Borasoy:2006qn}, it was shown that an on-site interaction such as those shown in
Eqs.~(\ref{contact:LO_a}) and~(\ref{contact:LO_b}) do not suffice to provide a favorable description of the $S$-wave phase shifts
except at very low momenta. Hence, smeared contact operators were introduced according to
\begin{equation}
\mathcal{O}_1^{(0)} \rightarrow \frac{1}{2L^3}  :\sum_{\vec{q}} f(\vec{q}\,)
\rho(\vec{q}\,)\rho(-\vec{q}\,):, 
\label{smear_a} 
\end{equation}
and
\begin{equation}
\mathcal{O}_2^{(0)} \rightarrow \frac{1}{2L^3}  :\sum_{\vec{q}} f(\vec{q}\,)
\rho_I^{}(\vec{q}\,)\rho_I^{}(-\vec{q}\,):,
\label{smear_b}
\end{equation}
where the smearing factor $f(\vec{q}\,)$ is
\begin{equation}
f(\vec{q}\,) \equiv f_0^{-1} \exp\left(-b_s^{} \frac{\vec q\,^4}{4}\right),
\label{smear_q4}
\end{equation}
with $b_s^{}$  a free parameter, and the normalization is given by
\begin{equation}
f_0^{} \equiv \frac{1}{L^3} \sum_{\vec{q}} \exp\left(-b_s^{} \frac{\vec q\,^4}{4}\right),
\end{equation}
with
\begin{eqnarray} 
\frac{\vec q\,^2}{2} &\equiv& \sum_{l=1}^3 \Bigg[\omega_0^{} - \omega_1^{} \cos\left(\frac{2\pi}{L} q_l^{}\right) 
+ \omega_2^{} \cos\left(\frac{4\pi}{L} q_l^{} \right) \nn \\
&& - \: \omega_3^{} \cos\left(\frac{6\pi}{L} q_l^{} \right)\Bigg], 
\label{q2smear}
\end{eqnarray}
where the $q_l^{}$ are lattice momentum components, and the $\mathcal{O}(a^4)$-improved 
hopping coefficients $\omega_i^{}$ are given in Table~\ref{hopping}.

In the analysis of the Ref.~\cite{Borasoy:2007vk}, smeared contact operators were found to dramatically improve the convergence
of the NLEFT expansion in the $S$-wave channels, at the price of introducing unwanted attractive forces in the
$P$-wave channels. By means of the projection operators~\cite{Epelbaum:2008vj}, 
\begin{eqnarray}
P^{(0,1)} &\equiv& \left(\frac{1}{4} - \frac{\vec{\sigma}_1 \cdot \vec{\sigma}_2}{4} \right) 
\left(\frac{3}{4} + \frac{\vec{\tau}_1 \cdot \vec{\tau}_2}{4} \right), \\
P^{(1,0)} &\equiv& \left(\frac{3}{4} + \frac{\vec{\sigma}_1 \cdot \vec{\sigma}_2}{4} \right)
\left(\frac{1}{4} - \frac{\vec{\tau}_1 \cdot \vec{\tau}_2}{4} \right),  
\end{eqnarray}
for the $(S, I) = (0, 1)$ and $(1, 0)$ channels, good agreement at LO in the $P$-wave channels can be recovered (although a similar
problem of unwanted forces in the $D$-wave channels persists). In the
present work, we use the corresponding smeared LO contact operators
\begin{eqnarray}
\mathcal{O}_{(0,1)}^{(0)} &\equiv& \frac{3}{32L^3} : \sum_{\vec{q}} f(\vec{q}\,)\rho(\vec{q}\,) \rho(-\vec{q}\,): \\
&& - \frac{3}{32L^3} :\sum_{\vec{q}} f(\vec{q}\,) \sum_{S}\rho_S^{}(\vec{q}\,) \rho_S^{}(-\vec{q}\,): \nn \\
&& + \frac{1}{32L^3} :\sum_{\vec{q}}f(\vec{q}\,) \sum_I\rho_I^{}(\vec{q}\,) \rho_I^{}(-\vec{q}\,) : \nn \\
&& - \frac{1}{32L^3} :\sum_{\vec{q}} f(\vec{q}\,)\sum_{S, I} \rho_{S,I}^{}(\vec{q}\,) \rho_{S,I}^{}(-\vec{q}\,):, \nn
\label{final_a}
\end{eqnarray}
for $(S, I) = (0, 1)$, and 
\begin{eqnarray}
\mathcal{O}_{(1,0)}^{(0)} &\equiv& \frac{3}{32L^3} :\sum_{\vec{q}} f(\vec{q}\,) \rho(\vec{q}\,) \rho(-\vec{q}\,) :\\
&&  + \frac{1}{32L^3} :\sum_{\vec{q}}f(\vec{q}\,) \sum_S \rho_S^{}(\vec{q}\,) \rho_S^{}(-\vec{q}\,): \nn \\
&& - \frac{3}{32L^3} : \sum_{\vec{q}} f(\vec{q}\,) \sum_I\rho_I^{}(\vec{q}\,) \rho_I^{}(-\vec{q}\,) : \nn \\
&& - \frac{1}{32L^3} : \sum_{\vec{q}} f(\vec{q}\,) \sum_{S,I} \rho_{S,I}^{}(\vec{q}\,) \rho_{S,I}^{}(-\vec{q}\,):, \nn 
\label{final_b}
\end{eqnarray}
for $(S, I) = (1, 0)$, where $\rho_{S}^{}(\vec{n})$ and $\rho_{S,I}^{}(\vec{n})$ are local spin density and local 
spin-isospin density operators, defined in App.~\ref{app_operators}. 

According to chiral EFT power counting, there are seven independent contact operators with 
two derivatives at NLO. Here, we use the basis and lattice formulation of Ref.~\cite{Borasoy:2007vi}, which leads
to the following NLO contact operators
\begin{eqnarray}
\mathcal{O}_1^{(2)} &\equiv& -\frac{1}{2} :\sum_{\vec n} \sum_{l} \rho(\vec{n}) \nabla_{l}^2 \rho(\vec{n}):,  \label{NLO_i}\\
\mathcal{O}_2^{(2)} &\equiv& -\frac{1}{2} :\sum_{\vec{n}} \sum_{I, l} \rho_I^{}(\vec{n}) \nabla_l^2 \rho_I^{}(\vec{n}):,  \\
\mathcal{O}_3^{(2)} &\equiv& -\frac{1}{2} :\sum_{\vec{n}} \sum_{S, l}\rho_S^{}(\vec{n}) \nabla_l^2\rho_S^{}(\vec{n}):, \\
\mathcal{O}_4^{(2)} &\equiv& -\frac{1}{2} :\sum_{\vec{n}}\sum_{S, I} \rho_{S, I}^{}(\vec{n}) 
\nabla_l^2 \rho_{S, I}^{}(\vec{n}):,  \label{projection_NLO} \\  
\mathcal{O}^{(2)}_5 &\equiv& \frac{1}{2} :\sum_{\vec{n}} \sum_{S} \nabla_S^{} \rho_{S}^{}(\vec{n}) 
\sum_{S^\prime} \nabla_{S^\prime}^{} \rho_{S^\prime}^{}(\vec{n}):, \\
\mathcal{O}^{(2)}_6 &\equiv& \frac{1}{2} :\sum_{\vec{n}} \sum_{S} \nabla_S^{} \rho_{S, I}^{}(\vec{n}) 
\sum_{S^\prime} \nabla_{S^\prime}^{} \rho_{S^\prime}^{} (\vec{n}):, \label{NLO_j} \\
\mathcal{O}^{(2)}_7 &\equiv& -\frac{i}{2} :\sum_{\vec{n}}\sum_{ l, S, l^\prime} \varepsilon_{l, S, l^\prime} 
\Bigg[\Pi_l^{}(\vec{n}) \nabla_{l^\prime}^{} \rho_S^{} (\vec{n}) \nn \\
&& + \: \Pi_{l, S}^{}(\vec{n}) \nabla_{l^\prime}^{}
\rho(\vec{n}) \Bigg]:,  
\end{eqnarray}
where $\Pi_l^{}(\vec{n})$ and $\Pi_{l, S}^{}(\vec{n})$ denote current density and spin-current density operators,
the lattice definitions of which are given in App.~\ref{app_operators}. Following the treatment of Ref.~\cite{Borasoy:2007vi} 
for the spin-orbit operator $\mathcal{O}_7^{(2)}$, we project onto $I=1$, giving
\begin{eqnarray}
\mathcal{O}_7^{(2)} &\rightarrow& -\frac{i}{2} \Bigg[\frac{3}{4}  :\sum_{\vec{n}} \sum_{l,S,l^\prime}
\varepsilon_{l,S,l^\prime}^{} \Big(\Pi_l^{}(\vec{n}) \nabla_{l^\prime}^{} \rho_S^{}(\vec{n}) \nn \\
&& \quad + \: \Pi_{l,S}^{}(\vec{n}) \nabla_{l^\prime}^{} \rho(\vec{n})\Big): \nonumber \\
&& + \: \frac{1}{4} :\sum_{\vec{n}} \sum_{l,S, l^\prime, I} \varepsilon_{l,S,l^\prime}^{}\Big(\Pi_{l, I}^{}(\vec{n}) \nabla_{l^\prime}^{} \rho_{S,I}^{}(\vec{n}) \nn \\
&& \quad + \: \Pi_{l,S,I}^{}(\vec{n})\nabla_{l^\prime}^{} \rho_I^{}(\vec{n})\Big):\Bigg], \label{NLO_7}
\end{eqnarray}
which eliminates lattice artifacts in the $S=1$ even-parity channels. For the derivative operator $\nabla_l^{}$ in the NLO contact terms, 
we use
\begin{equation}
\nabla_l^{} f(\vec{n}) \equiv \frac{1}{2a} \big[f(\vec{n} + a\hat e_l^{}) - f(\vec{n} - a\hat e_l^{})\big], 
\label{definition_nabla}
\end{equation} 
where $a$ is the spatial lattice spacing, and $\hat e_l^{}$ is a unit vector in spatial direction $l$. For the double derivative operator $\nabla_l^2$, 
we take 
\begin{equation}
\nabla_l^2 f(\vec{n}) \equiv \nabla_l^{} \big[\nabla_l^{} f(\vec{n})\big]. \label{definition_nabla2}
\end{equation}

In the radial transfer matrix formalism, we project each of the NLO contact operators onto the NN partial waves under consideration, 
such that $V_X^i$ is the matrix element of operator $i$ in channel $X$. If we denote the complete set of NLO contact interactions
by $V_{\mathrm{con}}^{(2)}$, we find
\begin{eqnarray}
\langle ^1S_0^{} |V_{\mathrm{con}}^{(2)}| ^1S_0^{} \rangle &=& \widetilde{C}_1^{} V_{{}^1S_0^{}}^1, \\
\langle ^1P_1^{} |V_{\mathrm{con}}^{(2)}| ^1P_1^{} \rangle &=& \widetilde{C}_4^{} V_{{}^1P_1^{}}^1, \label{projection_a}  \\
\langle ^3P_0^{} |V_{\mathrm{con}}^{(2)}| ^3P_0^{} \rangle &=& \widetilde{C}_5^{} V_{{}^3P_0^{}}^1 
+ \widetilde{C}_6^{} V_{{}^3P_0^{}}^5 + \widetilde{C}_7^{} V_{{}^3P_0^{}}^7, \\
\langle ^3P_1^{} |V_{\mathrm{con}}^{(2)}| ^3P_1^{} \rangle &=& \widetilde{C}_5^{} V_{{}^3P_1^{}}^1 
+ \widetilde{C}_6^{} V_{{}^3P_1^{}}^5 + \widetilde{C}_7^{} V_{{}^3P_1^{}}^7,
\end{eqnarray}
for the uncoupled channels, and 
\begin{eqnarray}
\langle {}^3SD_1^{} |V_{\mathrm{con}}^{(2)}| {}^3SD_1^{} \rangle
&=& \widetilde{C}_2 V_{{}^3SD_1}^1   + \widetilde{C}_3^{} V_{^3SD_1}^5, \\
\langle {}^3PF_2^{} |V_{\mathrm{con}}^{(2)}| {}^3PF_2^{} \rangle 
&=& \widetilde{C}_5^{} V_{{}^3PF_2}^1  + \widetilde{C}_6^{} V_{{}^3PF_2}^5  \nn \\
&& + \: \widetilde{C}_7^{} V_{{}^3PF_2}^7, \label{projection_b} 
\end{eqnarray}
for the coupled ones. It is clear that only certain combinations of the contact operators contribute to each
partial wave, which allows for a simplified fitting procedure. Specifically, we determine 
$C_{^1S_0^{}}^{}$ and $\widetilde{C}_1^{}$ by fitting the $^1S_0^{}$ channel, $\widetilde{C}_4^{}$ by means of the 
$^1P_1^{}$ channel, $\widetilde{C}_5^{}$, $\widetilde{C}_6^{}$ and $\widetilde{C}_7^{}$ from a simultaneous fit to the
$^3P_0^{}$, $^3P_1^{}$ and $^3P_2^{}$-$^3F_2^{}$ channels, and finally $C_{^3S_1^{}}^{}$, $\widetilde{C}_{2}^{}$ and 
$\widetilde{C}_3^{}$ by fitting the $^3S_1^{}$-$^3D_1^{}$ channel. 

We note that the 
fitted LECs $\widetilde{C}_i^{}$ are given in terms of those
of the NLO operators in Eqs.~(\ref{NLO_i}) through~(\ref{NLO_j}) and~(\ref{NLO_7}) by the relation
\begin{eqnarray}
\left[
\begin{array}{c}
\widetilde C_1 \\
\widetilde C_2 \\
\widetilde C_3 \\
\widetilde C_4 \\
\widetilde C_5 \\
\widetilde C_6 \\
\widetilde C_7
\end{array}
\right]
= 
\left[
\begin{array}{ccccccc}
1    &   1        &   $-3$   &   $-3$   &   $-1$   &   $-1$    &   0 \vspace{.07cm} \\
1    &   $-3$   &   1        &   $-3$   &   0        &      0       &   0 \vspace{.07cm} \\
0    &   0        &   0        &     0       &   1        &   $-3$    &  0 \vspace{.07cm} \\
1    &   $-3$   &   $-3$   &   9         &   $-1$  &     3       &   0 \vspace{.07cm} \\
1    &   1        &   1        &   1         &      0     &     0       &   0 \vspace{.07cm} \\
0    &   0        &   0        &     0       &     1      &     1       &   0 \vspace{.07cm} \\
0    &   0        &   0        &     0       &      0     &      0      &   1
\end{array}
\right]
\left[
\begin{array}{l}
C_{q^2} \vspace{.07cm} \\
C_{I^2,q^2} \vspace{.07cm} \\
C_{S^2,q^2} \vspace{.07cm} \\
C_{S^2,I^2,q^2} \vspace{.07cm} \\
C_{(q\cdot S)^2} \vspace{.07cm} \\
C_{I^2,(q\cdot S)^2} \vspace{.07cm} \\
C_{(q\times S) \cdot k}^{I=1}
\end{array}
\right],
\end{eqnarray}
which can be inverted in order to find the original LECs
$C_i^{}$, once the $\widetilde{C}_i^{}$ have been determined.


\subsection{Long-range interactions}

Next, we consider the long-range one-pion exchange (OPE) and two-pion exchange (TPE) contributions to the chiral EFT interaction.
The latter contributes at NLO and NNLO. At LO, the OPE potential is given by~\cite{Borasoy:2006qn,Lee:2008fa}
\begin{eqnarray}
V_{\mathrm{OPE}}^{(0)}(M_{\pi}^{}) &=& -\frac{g_A^2}{8 F_\pi^2}\sum_{S_1^{}, S_2^{}, I} \sum_{\vec{n}_1^{}, \vec{n}_2^{}} 
G_{S_1^{}, S_2^{}}^{} (\vec{n}_1^{} - \vec{n}_2^{}, M_\pi^{}) \nn \\
&\times& \rho_{S_1^{}, I}^{}(\vec{n}_1^{}) \rho_{S_2^{}, I}^{}(\vec{n}_2^{}),  
\end{eqnarray}
where the pion propagator is
\begin{eqnarray}
G_{S_1^{}, S_2^{}}^{}(\vec{n}_1^{} - \vec{n}_2^{}, M_\pi^{}) &\equiv& 
\frac{1}{L^3}\sum_{\vec{k}}\exp{\Bigg[i \frac{2\pi}{L} \vec{k} \cdot (\vec{n}_1^{} - \vec{n}_2^{})\Bigg]} \nn \\
&\times& G_{S_1^{}, S_2^{}}^{}(\vec k, M_\pi^{}),
\end{eqnarray}
with
\begin{equation}
G_{S_1^{}, S_2^{}}^{}(\vec k, M_\pi^{}) \equiv \frac{q_{S_1^{}}^{} q_{S_2^{}}^{}}{M_\pi^2 + \vec q\,^2},
\label{Gpi}
\end{equation}
where the $k_i^{}$ are lattice momentum components. For the denominator of
Eq.~(\ref{Gpi}), we take
\begin{eqnarray} 
\label{Eq:q2_OPE}
\vec q\,^2 &\equiv& 2 \sum_{l=1}^3 \Bigg[ \omega_0^{} - \omega_1^{} \cos\left(\frac{2\pi}{L} k_l^{}\right) 
+ \omega_2^{} \cos\left(\frac{4\pi}{L} k_l^{} \right) \nn \\
&& - \: \omega_3^{} \cos\left(\frac{6\pi}{L} k_l^{} \right)\Bigg], 
\end{eqnarray}
using the $\mathcal{O}(a^4)$-improved hopping coefficients $\omega_i^{}$ of Table~\ref{hopping}. 
For the numerator of Eq.~(\ref{Gpi}), we take
\begin{equation}
q_i^{} \equiv \sin\left(\frac{2\pi}{L}k_i^{}\right), 
\label{momentum:a}
\end{equation}
which coincides with the choice of derivative operator in Eq.~(\ref{definition_nabla}). We also include the 
isospin-breaking (IB) effects due to the pion mass differences. Specifically, we take
\begin{eqnarray}
V_{\mathrm{OPE}}^{(0)}(I=1) &=& 2V_{\mathrm{OPE}}^{(0)}(M_{\pi^{\pm}}^{}) - V_{\mathrm{OPE}}^{(0)}(M_{\pi^0}^{}), \\
V_{\mathrm{OPE}}^{(0)}(I=0) &=& -2 V_{\mathrm{OPE}}^{(0)}(M_{\pi^{\pm}}^{}) - V_{\mathrm{OPE}}^{(0)}(M_{\pi^0}^{}), 
\end{eqnarray}
for the isospin-triplet and isospin-singlet channels, respectively. This approach is consistent with the conventions of the 
Nijmegen partial wave analysis. For more details on the IB corrections to the NN interaction, see
Refs.~\cite{Epelbaum:2004fk,Epelbaum:2008ga,Machleidt:2011zz} (and references therein).

The first contribution from the TPE potential appears at NLO in chiral EFT. We note that several prior continuum calculations including TPE
exist. For instance, in Refs.~\cite{Friar:1994zz,Kaiser:1997mw}, dimensional regularization (DR) was used to remove the divergence appearing in the loop 
integral, and a non-local momentum-dependent form factor was applied to suppress the high-momentum contributions when solving the 
Lippmann-Schwinger equation. In Ref.~\cite{Epelbaum:2003gr}, another regularization called spectral function regularization (SFR) was proposed. 
Compared to DR, the SFR method introduces an additional cutoff to remove the short-range components of the TPE potential. Recently, 
a new position-space regularization was proposed in Refs.~\cite{Gezerlis:2013ipa,Gezerlis:2014zia,Epelbaum:2014efa}. The study of effects in nuclear
lattice EFT due to different choices of regularization of the TPE is beyond the scope of the current work.
In this work, we use the DR expressions with discretized lattice momenta. We also note that the lattice spacing serves as a natural UV cut-off.

Thus far, nuclear lattice EFT calculations have been performed with a lattice spacing of $a = 1.97$~fm, and hence the TPE potentials at NLO and NNLO
have not been included explicitly, but rather been absorbed into the contact terms. Since we are here studying the effects of reducing the lattice spacing
to $a \simeq 1$~fm, we shall for the first time include the full TPE structure. As for the smeared LO contact terms and the OPE potential, we define the 
lattice formulation of the TPE potential in momentum space, and Fourier transform the results to coordinate space. The TPE potential is of the form
\begin{eqnarray}
V_{\mathrm{TPE}}^{(2)} &=& \sum_{\vec{n}_1^{},\vec{n}_2^{}} \sum_{S_1^{},S_2^{}} 
T_{S_1^{}, S_2^{}}^{(2)}(\vec{n}_1^{}-\vec{n}_2^{})  \rho_{S_1^{}}^{}(\vec{n}_1^{})\rho_{S_2^{}}^{}(\vec{n}_2^{}) \nn \\
&+& \sum_{\vec{n}_1^{}, \vec{n}_2^{}} \sum_{I} 
W_C^{(2)}(\vec{n}_1^{} - \vec{n}_2^{}) \rho_I^{}(\vec{n}_1^{})\rho_I^{}(\vec{n}_2^{}) \nn \\
&+& \sum_{\vec{n}_1^{}, \vec{n}_2^{}} \sum_{S}
V_S^{(2)}(\vec{n}_1^{}-\vec{n}_2^{}) \rho_S^{}(\vec{n}_1^{})\rho_S^{}(\vec{n}_2^{}),
\label{NLO:tpep}
\end{eqnarray}
at NLO. The explicit expressions for the components of Eq.~(\ref{NLO:tpep}) are
\begin{eqnarray}
T_{S_1^{},S_2^{}}^{(2)} (\vec{n}_1^{} - \vec{n}_2^{}) &\equiv& 
\frac{1}{L^3}\sum_{\vec{k}}\exp{\Bigg[i \frac{2\pi}{L} \vec{k} \cdot (\vec{n}_1 - \vec{n}_2)\Bigg]} \nn \\
&\times& T_{S_1^{},S_2^{}}^{(2)} (\vec k),
\end{eqnarray}
with
\begin{equation}
T_{S_1^{},S_2^{}}^{(2)} (\vec k)
\equiv 18 g_A^4 F^{(2)}(\vec q\,) \, q_{S_1}^{} q_{S_2}^{},
\end{equation}
and
%
%
\begin{eqnarray}
W_{C}^{(2)} (\vec{n}_1^{} - \vec{n}_2^{}) &\equiv& 
\frac{1}{L^3}\sum_{\vec{k}}\exp{\Bigg[i \frac{2\pi}{L} \vec{k} \cdot (\vec{n}_1 - \vec{n}_2)\Bigg]} \nn \\
&\times& W_{C}^{(2)} (\vec k),
\end{eqnarray}
with
\begin{eqnarray}
W_{C}^{(2)} (\vec k) &\equiv& 
F^{(2)}(\vec q\,) \bigg[\frac{48 g_A^2 M_\pi^4}{4M_\pi^2 + \vec q\,^2}
+ 4M_\pi^2\left(5g_A^4 - 4g_A^2 -1 \right)  \nonumber \\
&& + \: \vec q\,^2 \( 23 g_A^4 - 10g_A^2 - 1 \) \bigg],
\end{eqnarray}
%
%
and
\begin{eqnarray}
V_{S}^{(2)} (\vec{n}_1^{} - \vec{n}_2^{}) &\equiv& 
\frac{1}{L^3}\sum_{\vec{k}}\exp{\Bigg[i \frac{2\pi}{L} \vec{k} \cdot (\vec{n}_1 - \vec{n}_2)\Bigg]} \nn \\
&\times& V_{S}^{(2)} (\vec k),
\end{eqnarray}
with
\begin{equation}
V_{S}^{(2)} (\vec k)
\equiv -18 g_A^4 F^{(2)}(\vec q\,) \, {\vec q}\,^2,
\end{equation}
%
%
where the function $F^{(2)}(\vec q\,)$ is given by
\begin{equation}
F^{(2)}(\vec q\,) \equiv -\frac{1}{768\pi^2 F_\pi^4}L(\vec q\,),
\end{equation}
and $L(\vec q\,)$ is the loop function 
\begin{equation}
L(\vec q\,) \equiv \frac{\sqrt{4 M_\pi^2 + \vec q\,^2}}{2|\vec q\,|} 
\log \left(\frac{\sqrt{4M_\pi^2 + \vec q\,^2} + |\vec q|}{\sqrt{4M_\pi^2 + \vec q\,^2} - |\vec q|}\right),
\end{equation}
in DR. In order to
coincide with the definitions of the derivative operator~(\ref{definition_nabla}) and the double-derivative 
operator~(\ref{definition_nabla2}), we take
\begin{equation}
q_i^{} \rightarrow \sin\left(\frac{2\pi}{L}k_i^{}\right),
\end{equation}
and 
\begin{equation}
q_i^2 \rightarrow \left[\sin\left(\frac{2\pi}{L}k_i^{}\right) \right]^2,
\label{momentum:b}
\end{equation}
which ensures that the divergences appearing in the loop diagrams can be absorbed by tuning the 
contact interaction LECs $C_i^{}$. 

Similarly, we parameterize the sub-leading (NNLO) contribution to the TPE as
\begin{eqnarray}
V_{\mathrm{TPE}}^{(3)} &=& \sum_{\vec{n}_1^{}, \vec{n}_2^{}} \sum_{S_1^{}, S_2^{}, I} 
T_{S_1^{}, S_2^{}}^{(3)}(\vec{n}_1^{}-\vec{n}_2^{})\rho_{S_1^{},I}^{}(\vec{n}_1^{})\rho_{S_2^{},I}^{}(\vec{n}_2^{}) \nn \\
&+& \sum_{\vec{n}_1^{}, \vec{n}_2^{}} \sum_{S,I} 
W_S^{(3)}(\vec{n}_1^{}-\vec{n}_2^{})\rho_{S,I}^{}(\vec{n}_1^{})\rho_{S,I}^{}(\vec{n}_2^{}) \nn \\
&+& \sum_{\vec{n}_1^{}, \vec{n}_2^{}} V_C^{(3)}(\vec{n}_1^{} - \vec{n}_2^{}) \rho(\vec{n}_1^{})\rho(\vec{n}_2^{}),
\end{eqnarray}
where
\begin{eqnarray}
T_{S_1^{},S_2^{}}^{(3)} (\vec{n}_1^{} - \vec{n}_2^{}) &\equiv& 
\frac{1}{L^3}\sum_{\vec{k}}\exp{\Bigg[i \frac{2\pi}{L} \vec{k} \cdot (\vec{n}_1 - \vec{n}_2)\Bigg]} \nn \\
&\times& T_{S_1^{},S_2^{}}^{(3)} (\vec k),
\end{eqnarray}
with
\begin{equation}
T_{S_1^{},S_2^{}}^{(3)} (\vec k)
\equiv c_4^{} F^{(3)}(\vec q\,) \, (4 M_\pi^2 + \vec{q}\,^2) \, q_{S_1}^{} q_{S_2}^{},
\end{equation}
%
%
%
and
\begin{eqnarray}
W_{S}^{(3)} (\vec{n}_1^{} - \vec{n}_2^{}) &\equiv& 
\frac{1}{L^3}\sum_{\vec{k}}\exp{\Bigg[i \frac{2\pi}{L} \vec{k} \cdot (\vec{n}_1 - \vec{n}_2)\Bigg]} \nn \\
&\times& W_{S}^{(3)} (\vec k),
\end{eqnarray}
with
\begin{equation}
W_{S}^{(3)} (\vec k)
\equiv -c_4^{} F^{(3)}(\vec q\,) \, \vec{q}\,^2,
\end{equation}
%
and 
\begin{eqnarray}
V_{C}^{(3)} (\vec{n}_1^{} - \vec{n}_2^{}) &\equiv& 
\frac{1}{L^3}\sum_{\vec{k}}\exp{\Bigg[i \frac{2\pi}{L} \vec{k} \cdot (\vec{n}_1 - \vec{n}_2)\Bigg]} \nn \\
&\times& V_{C}^{(3)} (\vec k),
\end{eqnarray}
with
\begin{equation}
V_{C}^{(3)} (\vec k)
\equiv 6 F^{(3)}(\vec q\,) \, (2 M_\pi^2 + \vec{q\,}^2)
\left[2M_\pi^2 (2 c_1^{} - c_3^{}) - c_3^{} \vec{q \,}^2 \right],
\end{equation}
%
where the function $F^{(3)}(\vec q\,)$ is given by
\begin{equation}
F^{(3)}(\vec q\,) \equiv -\frac{g_A^2}{64\pi f_{\pi}^4} A(\vec q\,),
\end{equation}
and $A(\vec q\,)$ is the loop function 
\begin{equation}
A(\vec q\,) \equiv \frac{1}{2|\vec q\,|} \arctan\left({\frac{|\vec q\,|}{2M_\pi^{}}}\right),
\end{equation}
in DR. For the momenta $\vec q$, we again apply the conventions of 
Eqs.~(\ref{momentum:a}) and~(\ref{momentum:b}).


\section{Results \label{results}}

We now turn to a description of our calculational methods.
We take $F_\pi^{} = 92.2$~MeV for the pion decay constant, and $g_A^{} = 1.29$ for the 
nucleon axial coupling constant to account for the Goldberger-Treiman discrepancy~\cite{Epelbaum:2004fk}.
For the nucleon mass, we use $m_N^{} = 938.38$~MeV, and for the charged and neutral pion masses, we take
$M_{\pi^\pm}^{} = 139.75$~MeV and $M_{\pi^0}^{} = 134.98$ MeV, respectively. We use the isospin-averaged pion mass
\begin{equation}
M_\pi^{} \equiv \frac{2}{3} M_{\pi^\pm}^{} + \frac{1}{3} M_{\pi^0}^{} = 138.03~\mathrm{MeV},
\end{equation}
in the TPEP expressions at NLO and NNLO. For the constants $c_1^{}$, $c_3^{}$ and $c_4^{}$ that appear in the 
TPEP at NNLO, we use $c_1^{} = -1.10(3)$ GeV$^{-1}$, $c_3^{} = -5.54(6)$ GeV$^{-1}$ and 
$c_4^{} = 4.17(4)$ GeV$^{-1}$ from the accurate Roy-Steiner analysis of pion-nucleon scattering
adopted to the counting of the nucleon mass used here~\cite{Hoferichter:2015tha}. Also, as the
uncertainties of these LECs are very small, we only consider the central values in the following. 


\begin{table}[t]
\begin{center}
\caption{Summary of lattice spacings $a$ (spatial) and $a_t^{}$ (temporal) and box dimensions $L$.  
The physical spatial lattice volume $V$ is kept constant at $(La)^3 \simeq (63~\mathrm{fm})^3$. 
\label{lattice_spacing}}
\smallskip
\begin{tabular*}{0.475\textwidth}{@{\extracolsep{\fill}}ccccc}
\hline\hline
\noalign{\smallskip}
 $a^{-1}$~[MeV]   & $a_t^{-1}$~[MeV]  & $a$~[fm]     &     $L$ &      $La$~[fm]   
 \smallskip \\
 \hline
 100  &  150	&     1.97      &     32            &     63.14   \\
 120  &  216  	&	1.64      &     38            &     62.48   \\
 150  &  337.5 	&	1.32      &     48            &     63.14    \\
 200  &  600	&    0.98      &     64            &     63.14   \\
\hline\hline
\end{tabular*}
\end{center}
\end{table}


\begin{table}[t]
\begin{center}
\caption{Summary of the fitting procedure, indicating which parameters are fitted to what scattering channel 
at each order in NLEFT, and the resulting $\chi^2/N_\mathrm{dof}^{}$ (for $a=0.98$~fm). 
\label{fit:procedure}}
\smallskip
\begin{tabular*}{0.475\textwidth}{@{\extracolsep{\fill}}lllc}
\hline \hline
\noalign{\smallskip}
 order                          &           fit channels           &      fit parameters                                     &  $\chi^2/N_\mathrm{dof}^{}$ 
  \smallskip \\
\hline
\noalign{\smallskip}
LO & $^1S_0^{}$,  $^3S_1^{}$  &    $C_{^1S_0}^{}$, $C_{^3S_1^{}}^{}$, $b_s^{}$   &    $30.38$ \\  
 \hline
 \multirow{4}{*}{NLO}    &  $^1S_0^{}$                     &    $C_{^1S_0}^{}$, $\widetilde{C}_1^{}$     &     $1.77$                        \\
                                     &  $^3S_1^{}$, $\epsilon_1^{}$ & $C_{^3S_1}^{}$, $\widetilde{C}_2^{}$, $\widetilde{C}_3^{}$ &  $88.81$         \\
                                     &  $^1P_1^{}$                      &    $\widetilde{C}_4^{}$                            &     $11.94$                                 \\
                                     & $^3P_0^{}$, $^3P_1^{}$, $^3P_2^{}$ & $\widetilde{C}_5^{}$, $\widetilde{C}_6^{}$, $\widetilde{C}_7^{}$ &  $6.51$  \\
 \hline
  \multirow{4}{*}{NNLO}    &  $^1S_0^{}$                     &    $C_{^1S_0}^{}$, $\widetilde{C}_1^{}$    &     $0.36$                          \\
                                     &  $^3S_1^{}$, $\epsilon_1^{}$ & $C_{^3S_1}^{}$, $\widetilde{C}_2^{}$, $\widetilde{C}_3^{}$ &    $28.81$      \\
                                     &  $^1P_1^{}$                      &    $\widetilde{C}_4$                            &     $2.79$                           \\
                                     & $^3P_0^{}$, $^3P_1^{}$, $^3P_2^{}$ & $\widetilde{C}_5^{}$, $\widetilde{C}_6^{}$, $\widetilde{C}_7^{}$ &   $25.59$ \\
\hline\hline
\end{tabular*}
\end{center}
\end{table}


\begin{table}[t]
\begin{center}
\caption{Fitted constants and low-energy parameters for $a = 0.98$~fm. The LO constants $C_{^1S_0}^{}$ and 
$C_{^3S_1}^{}$ are given in units of [$10^{-4}$ MeV$^{-2}$], and the $C_i^{}$ of the NLO interaction in units of [$10^{-8}$ MeV$^{-4}$]. 
Due to the large lattice ($L = 64$) for $a = 0.98$~fm, an uncertainty analysis using the variance-covariance matrix as
in Table~\ref{LECs:uncertainty} was numerically unfeasible. Hence, an estimated uncertainty of $2\%$ has been 
assigned, which is consistent with the uncertainties for larger $a$. 
For entries with a dagger~($\dagger$), the deuteron energy $E_d^{}$ has been included as an additional constraint.
\label{Swave_parameter_98}}
\smallskip
\begin{tabular*}{0.475\textwidth}{@{\extracolsep{\fill}} lrrr}
\hline\hline
\noalign{\smallskip} 
& \multicolumn{1}{c}{LO} & \multicolumn{1}{c}{NLO} & \multicolumn{1}{c}{NNLO} \smallskip \\
\hline
$C_{^1S_0}^{}$  				&  $-0.101(2)$   & $-0.105(2)$ & $-0.106(2)$ \\
$C_{^3S_1}^{}$  				&  $-0.118(2)$   & $-0.087(2)$ & $-0.088(2)$ \\
$b_s^{}$             				&  $0.399(8)$    & $-$  & $-$ \\
$C_{q^2}^{}$      				& $-$ & $0.00440(8)$  & $0.135(2)$                    \\
$C_{I^2,q^2}^{}$    				& $-$ & $0.0373(8)$   & $0.0303(6)$                     \\ 
$C_{S^2,q^2}^{}$       			& $-$ & $-0.0292(6)$  & $-0.0301(6)$                   \\ 
$C_{S^2,I^2,q^2}^{}$     			& $-$ & $-0.00190(4)$  & $-0.0254(5)$                  \\ 
$C_{(q\cdot S)^2}^{}$       			& $-$ & $0.0378(8)$  & $0.0360(7)$                   \\ 
$C_{I^2,(q\cdot S)^2}^{}$       		& $-$ & $0.00200(4)$  & $0.0212(4)$                   \\ 
$C_{(q\times S) \cdot k}^{I=1}$ 		& $-$ & $0.0150(3)$  & $0.0165(3)$                  \\  
\hline
$E_d^{}$ [MeV]            			& $2.16(4)$     & $2.22(4)^\dagger$   & $2.22(4)^\dagger$     \\
$r_{^1S_0}^{}$ [fm]  				& $2.12(4)$     & $2.50(5)$   & $2.63(5)$      \\
$a_{^1S_0}^{}$ [fm]   			& $-22.5(4)$    & $-23.4(5)$& $-23.7(5)$  \\
$r_{^3S_1}^{}$ [fm]  				& $1.73(3)$     & $1.70(3)$   &  $1.74(3)$     \\
$a_{^3S_1}^{}$ [fm]   			& $5.4(1)$       & $5.4(1)$   &  $5.4(1)$    \\ [2pt]
\hline\hline
\end{tabular*}
\end{center}
\end{table}


\begin{table*}[t]
\begin{center}
\caption{Fitted constants and low-energy $S$-wave parameters for $a = 0.98$~fm. The LO constants $C_{^1S_0}^{}$ and 
$C_{^3S_1}^{}$ are given in units of [$10^{-4}$ MeV$^{-2}$], and the $C_i^{}$ of the NLO interaction in units of [$10^{-8}$ MeV$^{-4}$].
The smearing parameter $b_s^{}$ of the LO contact interactions is determined by the LO fit, and thereafter kept fixed at NLO and NNLO. 
The values in parentheses are the uncertainties calculated using the variance-covariance matrix according to Eq.~(\ref{uncertainty_coefficient}). 
\label{LECs:uncertainty}}
\smallskip
\begin{tabular*}{\textwidth}{@{\extracolsep{\fill}}llrrr}
\hline\hline
\noalign{\smallskip}
order         &  fit parameters  &  $a = 1.97$~fm & $a = 1.64$~fm & $a = 1.32$~fm \smallskip \\
\hline
\multirow{3}{*}{LO}    & $C_{^1S_0}^{}$ 	&$-0.4676(2)$ & $-0.3290(7)$ & $-0.201(5)$  \\
                                   & $C_{^3S_1}^{}$	&$-0.6377(2)$ & $-0.4482(2)$&  $-0.265(5)$  \\
                                   & $b_s^{}$          	&$0.0524(2)$  & $0.0917(2)$ &  $0.173(6)$ \\
\hline
\multirow{9}{*}{NLO}  & $C_{^1S_0}^{}$                                 & $-0.5(1) $ & $-0.35(2)$ &  $ -0.220(2)$  \\
                                  & $C_{^3S_1}^{}$                                 & $-0.44(7)$ & $-0.21(1)$ &  $ -0.152(4)$  \\
                                  & $C_{q^2}^{}$                                     & $-0.05(3)$  & $-0.032(9)$ & $-0.006(1)$  \\
                                  & $C_{I^2,q^2}^{}$                               & $ 0.08(2)$  & $0.075(2)$  &  $ 0.052(1)$   \\
                                  & $C_{S^2,q^2}^{}$                             & $-0.06(3)$  & $-0.046(3)$ &  $-0.0341(7)$  \\
                                  & $C_{S^2,I^2,q^2}^{}$                       & $ 0.03(2)$   & $0.029(2)$  &  $0.0081(2)$  \\
                                  & $C_{(q\cdot S)^2}^{}$                      & $ 0.11(2)$   &  $ 0.091(4)$ &  $0.0553(2)$  \\
                                  & $C_{I^2,(q\cdot S)^2}^{}$                & $-0.11(2)$   &  $-0.074(4)$ &  $-0.0240(8)$  \\
                                  & $C_{(q\times S) \cdot k}^{I=1}$   & $0.037(8)$  &  $ 0.026(4)$ &  $0.019(2)$ \\
\hline
\multirow{9}{*}{NNLO}  & $C_{^1S_0}^{}$  & $-0.5(1)$& $-0.33(4)$ & $-0.21(2)$     \\
                                     & $C_{^3S_1}^{}$  & $-0.5(1)$& $-0.22(1)$ & $-0.15(2)$    \\
                  		   & $C_{q^2}^{}$       & $ 0.08(3)$& $ 0.093(7)$  & $0.118(7)$    \\
                  	            &  $C_{I^2,q^2}^{}$    & $ 0.07(2)$& $ 0.0668(4)$   & $0.045(4)$  \\
	                             & $C_{S^2,q^2}^{}$    & $-0.06(3)$& $-0.05(2)$  & $-0.036(7)$     \\
                                     & $C_{S^2,I^2,q^2}^{}$  & $ 0.01(2)$ & $0.005(3)$   & $-0.014(4)$   \\
                 		   & $C_{(q\cdot S)^2}^{}$  & $ 0.10(3)$ & $0.086(7)$   & $0.056(4)$     \\
                  		   & $C_{I^2,(q\cdot S)^2}^{}$   & $-0.10(3)$ & $-0.055(4)$  & $-0.006(4)$     \\
                                     & $C_{(q\times S) \cdot k}^{I=1}$   & $0.031(8)$  & $0.025(4)$  & $0.018(2)$ \\
\hline\hline                          
\end{tabular*} 
\end{center}
\end{table*}


\begin{table*}[t]
\begin{center}
\caption{Low-energy $S$-wave parameters, as a function of the lattice spacing $a$ and the order of the NLEFT expansion.
$E_d^{}$ is the deuteron binding energy, and the $a_i^{}$ and $r_i^{}$ denote the scattering lengths and effective ranges in channel $i$.
The experimental value of $E_d^{}$ is from Ref.~\cite{VanDerLeun:1982bhg}, and the scattering lengths and effective ranges are from 
Ref.~\cite{Machleidt:2000ge}. For entries marked with a dagger~($\dagger$), 
the empirical deuteron energy $E_d^{}$ has been included in the fit as an additional constraint.
\label{Swave:parameter}}
\smallskip
\begin{tabular*}{\textwidth}{@{\extracolsep{\fill}}lllllll}
\hline\hline
\noalign{\smallskip}
order         &   $a$ [fm]   & $E_d^{}$~[MeV]  & $r_{^1S_0}^{}$~[fm]  & $a_{^1S_0}^{}$~[fm]   & $r_{^3S_1}^{}$~[fm] & $a_{^3S_1}^{}$~[fm]  
\smallskip \\          
\hline
 \noalign{\smallskip}
 \multirow{3}{*}{LO}&   $1.97$     & $2.00(1)$    & $2.041(1)$            & $-22.4(4)$            &  $1.686(1)$           &  $5.46(1)$  \\
                               &   $1.64$     & $2.07(1)$    & $2.093(5)$            & $-22.5(7)$            &  $1.6932(8)$         &  $5.45(1)$  \\
                               &   $1.32$     & $2.12(2)$    & $2.11(2)$              & $-22.5(5)$            &  $1.71(1)$             &  $5.44(1)$  \\ 
 \hline 
 \multirow{3}{*}{NLO}& $1.97$    &  $2.2246(3)^\dagger$  & $2.4(6)$             & $-23(4)$               & $1.79(3)$              &  $5.31(2)$   \\
                                 & $1.64$     & $2.2246(1)^\dagger$  & $2.3(1)$             & $-23(2)$               & $1.73(1)$              &  $5.33(1)$   \\
                                 & $1.32$     & $2.2246(1)^\dagger$  & $2.47(3)$           & $-23(1)$               & $1.70(1)$              &  $5.336(9)$   \\
 \hline
 \multirow{3}{*}{NNLO}& $1.97$  & $2.2246(3)^\dagger$   & $2.6(6)$            & $-24(4)$               & $1.82(3)$              & $5.35(2)$    \\
                                    & $1.64$  & $2.2246(1)^\dagger$   & $2.5(3)$            & $-23(2)$               & $1.74(1)$              & $5.36(1)$    \\
                                    & $1.32$  & $2.22457(7)^\dagger$  & $2.6(2)$          &  $-23(1)$               & $1.744(7)$            & $5.382(5)$    \\
 \hline 
 \noalign{\smallskip}      
experiment    &  $-$   &     $2.224575(9)$ &  $2.77(5)$      &  $-23.740(20)$ & $1.753(8)$  & $5.419(7)$   
\smallskip \\ 
\hline\hline 
\end{tabular*}
\end{center} 
\end{table*}
 

\begin{figure*}[t]
\begin{center}
\includegraphics[width=\textwidth]{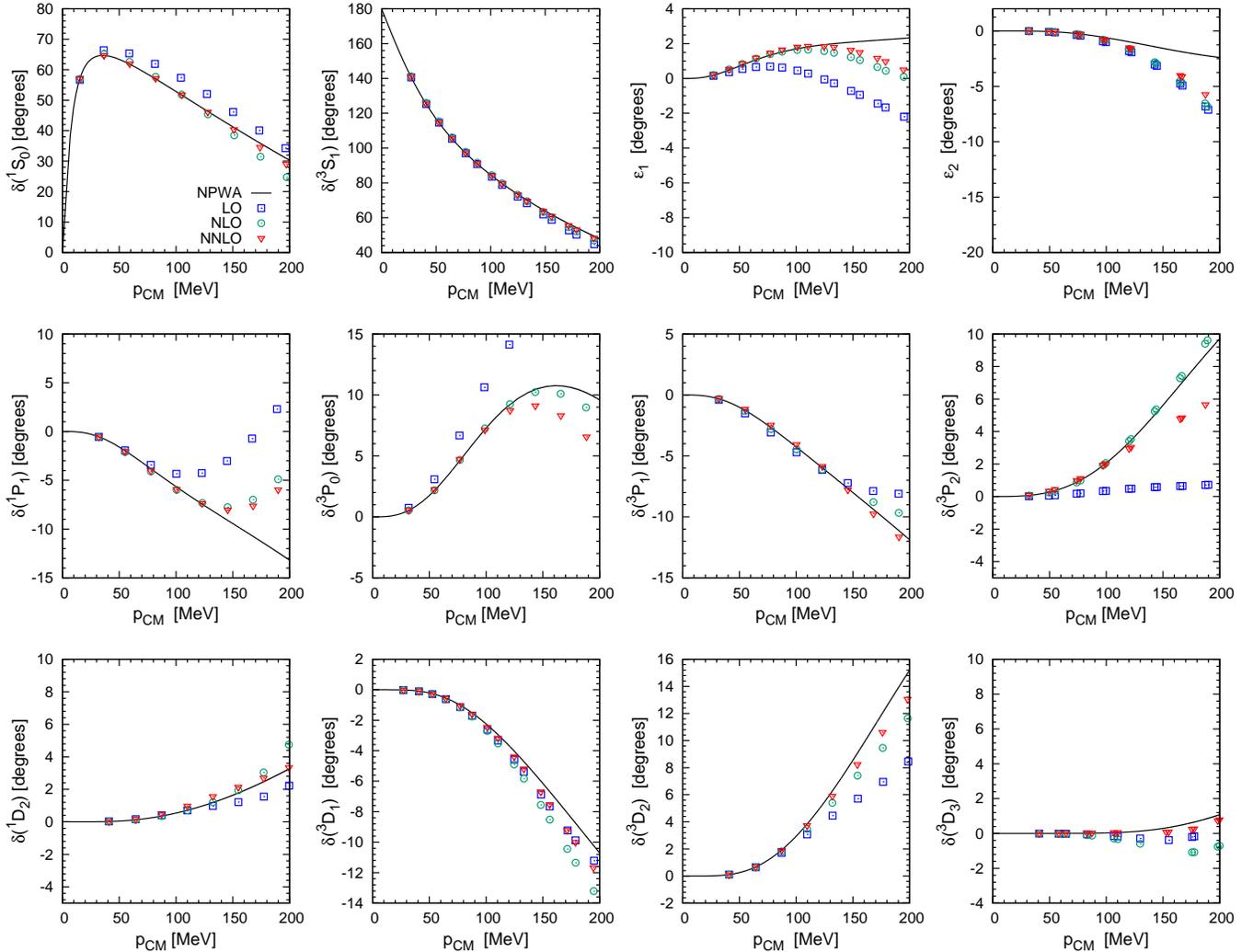}
\caption{(color online). Phase shifts and mixing angles for neutron-proton scattering up to NNLO in NLEFT, 
for our smallest (spatial) lattice spacing of $a = 0.98$~fm~$ = (200~\mathrm{MeV})^{-1}$ and a temporal
lattice spacing $a_t^{} = (600~\mathrm{MeV})^{-1}$. The (blue) squares, (green) circles and (red) triangles 
denote LO, NLO and NNLO results, respectively. The Nijmegen PWA is shown by the solid black line.
\label{lattice_spacing_98}}
\end{center}
\end{figure*}


\begin{figure*}[t]
\begin{center}
\includegraphics[width=\textwidth]{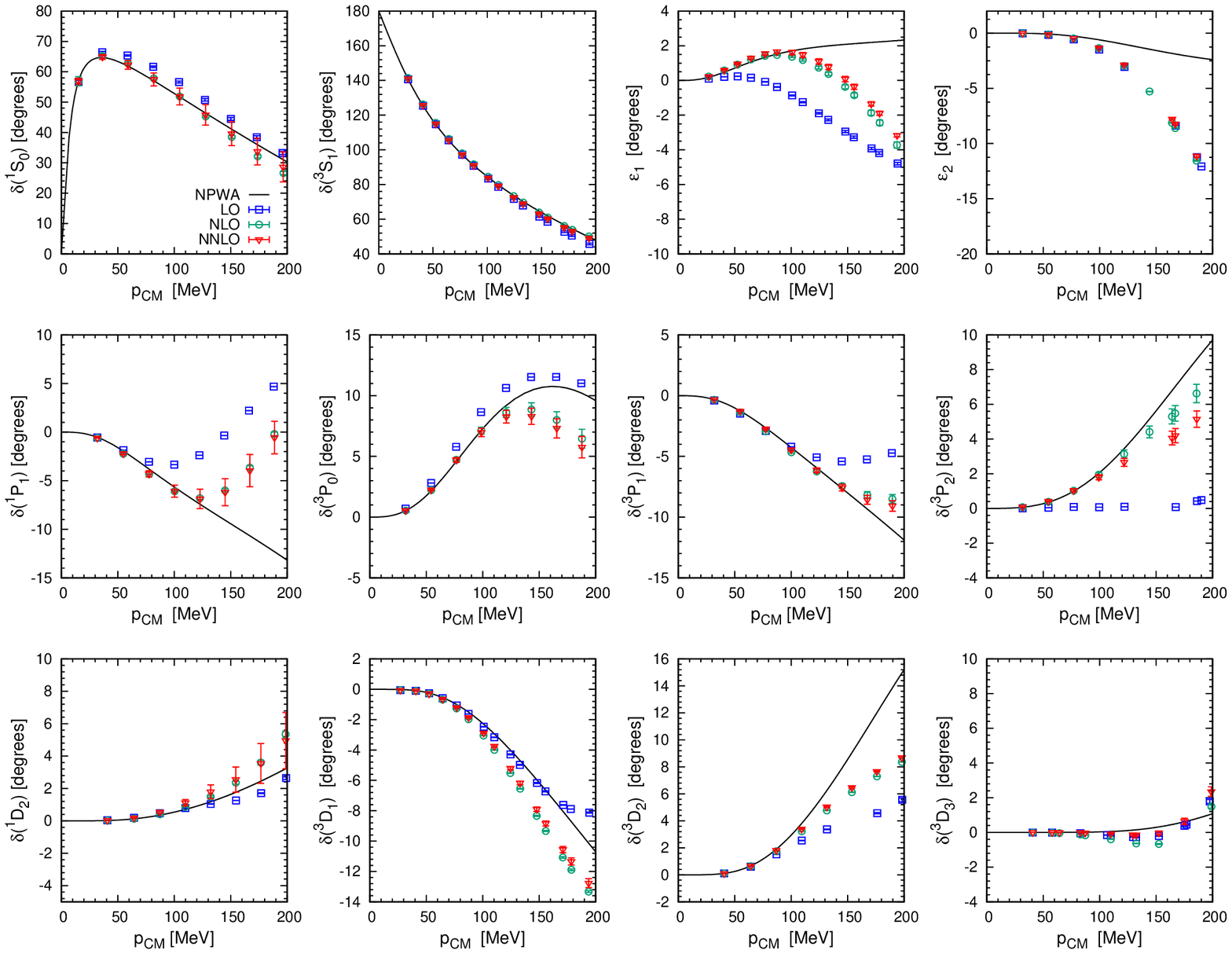}
\caption{(color online). Phase shifts and mixing angles for neutron-proton scattering up to NNLO in NLEFT, 
for $a = 1.32$~fm~$ = (150~\mathrm{MeV})^{-1}$. For notations, see  Fig.~\ref{lattice_spacing_98}.
\label{lattice_spacing_132}}
\end{center}
\end{figure*}


\begin{figure*}[t]
\begin{center}
\includegraphics[width=\textwidth]{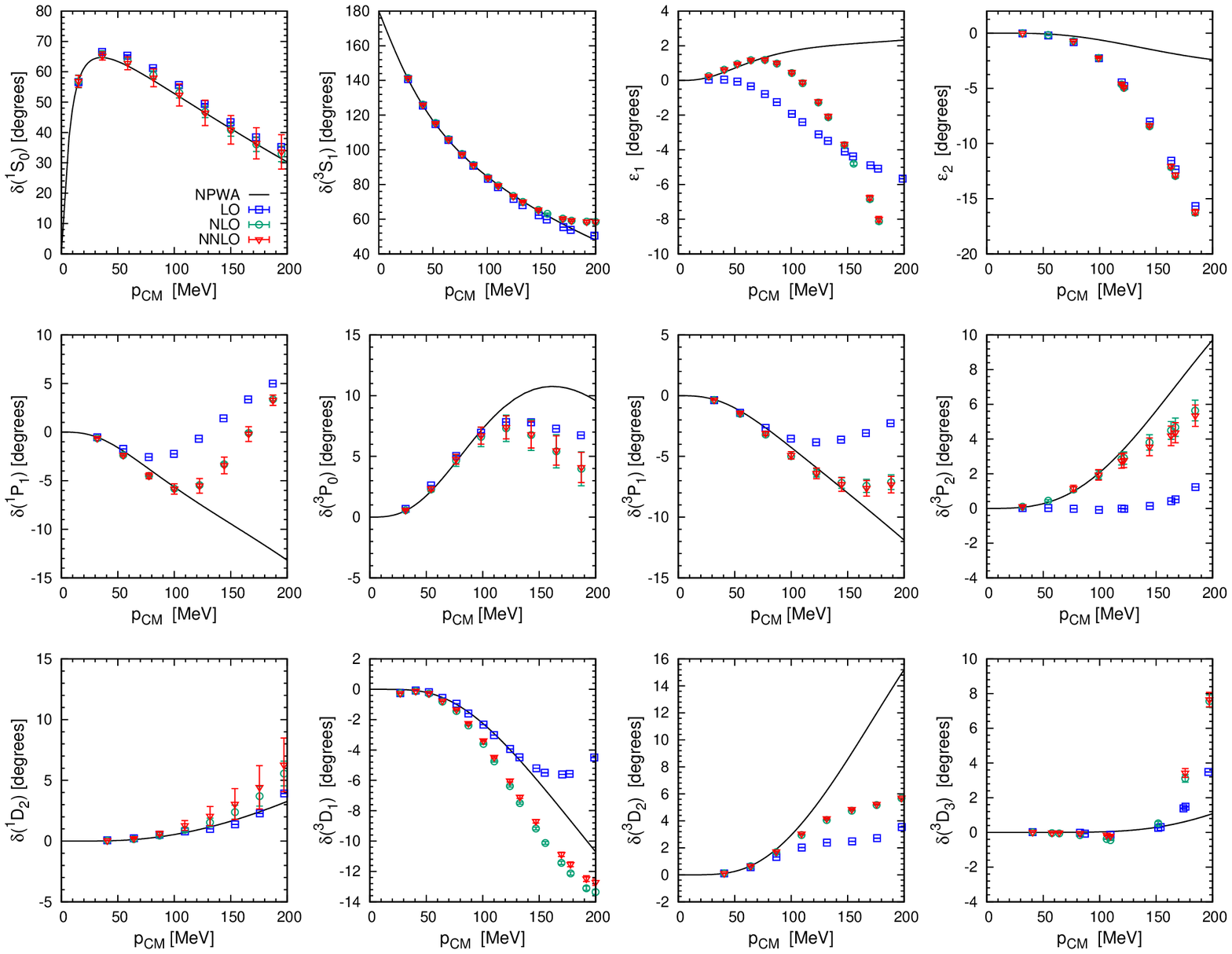}
\caption{(color online). Phase shifts and mixing angles for neutron-proton scattering up to NNLO in NLEFT, 
for $a = 1.64$~fm~$ = (120~\mathrm{MeV})^{-1}$. For notations, see  Fig.~\ref{lattice_spacing_98}.
\label{lattice_spacing_164}}
\end{center}
\end{figure*}


\begin{figure*}[t]
\begin{center}
\includegraphics[width=\textwidth]{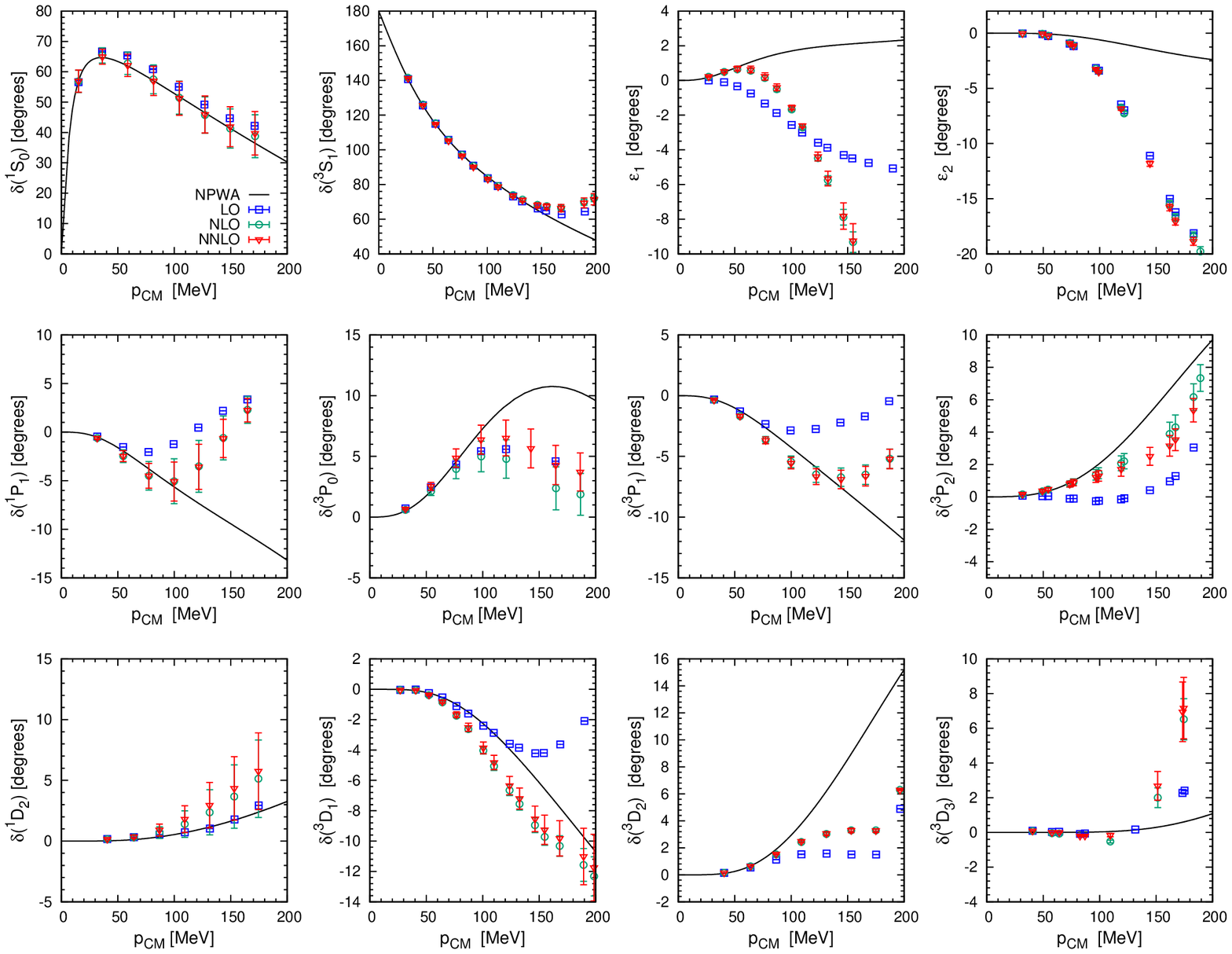}
\caption{(color online). Phase shifts and mixing angles for neutron-proton scattering up to NNLO in NLEFT, 
for a (spatial) lattice spacing $a = 1.97$~fm~$ = (100~\mathrm{MeV})^{-1}$ and a temporal
lattice spacing $a_t^{} = (150~\mathrm{MeV})^{-1}$. For notations, see Fig.~\ref{lattice_spacing_98}.
\label{lattice_spacing_197}}
\end{center}
\end{figure*}


We determine the optimal parameter values for the NLEFT action up to NNLO by performing a chi-square fit to neutron-proton phase shifts and mixing angles.
For this purpose, we define the uncertainties of the empirical scattering observables (in each partial wave) according to 
Refs.~\cite{Epelbaum:2014efa,Epelbaum:2014sza}, which gives
\begin{eqnarray}
\Delta_i^{} &\equiv& \max\Bigg[\Delta_i^{\mathrm{PWA}}, \bigg|\delta_i^{\mathrm{NijmI}} - \delta_i^{\mathrm{PWA}}\bigg|, 
\label{phase_error} \\
&& \quad \bigg|\delta_i^{\mathrm{NijmII}} - \delta_i^{\mathrm{PWA}}\bigg|, 
\bigg|\delta_i^{\mathrm{Reid93}} -\delta_i^{\mathrm{PWA}}\bigg|\Bigg], \nonumber
\end{eqnarray}
where $\Delta_i^{\mathrm{PWA}}$ denotes the uncertainty of the PWA, while $\delta_i^{\mathrm{PWA}}$ signifies 
the phase shift (or mixing angle) in channel $i$ of the PWA (see also Ref.~\cite{Stoks:1993tb}). Furthermore, 
$\delta_i^{\mathrm{NijmI}}$, $\delta_i^{\mathrm{NijmII}}$ and $\delta_i^{\mathrm{Reid93}}$ refer to the PWA results 
based on the Nijmegen~I, Nijmegen~II and Reid93~NN potentials, respectively. Hence, a measure of systematical error in the PWA
is accounted for in our analysis. The $\chi^2$ function to be minimized is defined as 
\begin{eqnarray}
\chi^2 \equiv \sum_i \frac{\left(\delta_i^{\mathrm{PWA}} - \delta_i^{\mathrm{cal}}\right)^2}{\Delta_i^2},  
\label{chisquare}
\end{eqnarray}
where $i$ runs over all values of $p_\mathrm{CM}^{}$ and channels included in the analysis. In Eq.~(\ref{chisquare}),
$\delta_i^{\mathrm{PWA}}$ is the phase shift (or mixing angle) at a given momentum $p_\mathrm{CM}^{}$ from the Nijmegen PWA, 
$\delta_i^{\mathrm{cal}}$ is the corresponding calculated NLEFT value, and $\Delta_i^{}$ is given by Eq.~(\ref{phase_error}).

When fitting the phase shifts and mixing angles of the Nijmegen partial wave analysis, we note certain simplifying features.
Specifically, at LO we determine $C_{^1S_0}^{}$, $C_{^3S_1}^{}$, and the smearing parameter $b_s^{}$, by fitting 
the $^1S_0$ and $^3S_1$ phase shifts.  At NLO and NNLO, we no longer update the value of $b_s^{}$. 
At NLO, we determine $C_{^1S_0}^{}$ and $\widetilde{C}_1^{}$ by fitting the $^1S_0$ phase shift, $C_{^3S_1}^{}$, $\widetilde{C}_2^{}$ 
and $\widetilde{C}_3^{}$ by fitting the $^3S_1$ phase shift and the mixing angle $\epsilon_1^{}$, $\widetilde{C}_4^{}$ by 
fitting the $^1P_1$ phase shift, and finally $\widetilde{C}_5^{}$, $\widetilde{C}_6^{}$ and $\widetilde{C}_7^{}$ by fitting the 
the $^3P_0$, $^3P_1$ and $^3P_2$ phase shifts. The NNLO fits are similar, apart from the inclusion of the NNLO TPEP operators.
We do not take the deuteron binding energy $E_d^{}$ as an additional constraint in the LO fits, as we do not expect $E_d^{}$ to be
accurately reproduced in an LO calculation. At NLO and NNLO, the experimental value $E_d = 2.224575(9)$ MeV is taken as 
an additional constraint. At LO, we fit up to center-of-mass momenta of 
$p_{\mathrm{CM}}^{\mathrm{max}} = 100$~MeV, while at NLO and NNLO we fit up to $p_{\mathrm{CM}}^{\mathrm{max}} = 150$~MeV. 
Our fitting procedure at each order in NLEFT is summarized in Table~\ref{fit:procedure}.

\subsection{Phase shifts and mixing angles to NNLO} 

Prior NLEFT work has used a relatively coarse lattice spacing of $a = 1.97$~fm, which
corresponds to a momentum cutoff $\Lambda \sim \pi/a = 314$~MeV. This relatively low cutoff may induce significant lattice
artifacts, particularly at high momenta. With this in mind, we here aim to study the NN scattering problem for
$a = (200~\mathrm{MeV})^{-1} = 0.98$~fm, with a temporal lattice spacing of $a_t^{} = (600~\mathrm{MeV})^{-1}$. The number of
lattice points in each spatial dimension is $L = 64$, thus the physical volume is $V = (La)^3 \simeq (63~\mathrm{fm})^3$, which is 
expected to be large enough to accommodate the NN system without introducing significant finite volume effects for the energy region 
$p_{\mathrm{CM}}^{} < 200$~MeV studied here. Our lattice parameters are summarized in Table~\ref{lattice_spacing}.

First, we consider the problem of neutron-proton scattering by treating all orders in NLEFT up to NNLO non-perturbatively,
similar to what is done in the continuum. This means
that we construct the transfer matrix according to
\begin{equation}
\label{Eq:Transfer_matrix_non-perturbative}
M \equiv \: :\exp\big[ -\alpha_t^{} (H_\mathrm{free}^{} + V_\mathrm{LO}^{} + V_\mathrm{NLO}^{} 
+ V_\mathrm{NNLO}^{}) \big]:,
\end{equation}
where the potential terms are given by
\begin{equation}
V_{\mathrm{LO}}^{} = C_{^1S_0}^{} \mathcal{O}_{(0,1)}^{(0)} + C_{^3S_1}^{} \mathcal{O}_{(1,0)}^{(0)} 
+ V_{\mathrm{OPE}}^{(0)},
\end{equation}
at LO,
\begin{eqnarray}
V_{\mathrm{NLO}}^{} &=& 
C_{q^2}^{} \mathcal{O}_{1}^{(2)} 
+ C_{I^2, q^2}^{} \mathcal{O}_{2}^{(2)} 
+ C_{S^2, q^2}^{} \mathcal{O}_{3}^{(2)} \nn \\
&& + \: C_{S^2, I^2, q^2}^{} \mathcal{O}_{4}^{(2)} 
+ C_{(q\cdot S)^2}^{} \mathcal{O}_{5}^{(2)} \nn \\
&& + \: C_{I^2, (q\cdot S)^2}^{} \mathcal{O}_{6}^{(2)} 
+ C_{(q\times S)\cdot k}^{I=1} \mathcal{O}_{7}^{(2)} \nn \\
&& + \: V_{\mathrm{TPE}}^{(2)} 
\label{Eq:V_NLO}, 
\end{eqnarray}
at NLO, and
\begin{equation}
V_{\mathrm{NNLO}}^{} =  V_{\mathrm{TPE}}^{(3)},
\end{equation}
at NNLO.
Our results for the smallest lattice spacing, $a = 0.98$~fm, are shown in Fig.~\ref{lattice_spacing_98}. 
Clearly, the description of the 
$S$-wave channels is quite good even at LO, particularly for $^3S_1$. Compared to LO, significant improvements occur
at NLO and NNLO, in particular for the $^1P_1$, ${}^3P_0$ and $^3P_2$ channels, as well as for the mixing angle $\epsilon_1^{}$. 
While the NLO contributions appear central for a good description of the $P$-waves and $\epsilon_1^{}$, the TPE contributions at NNLO do not
appear to produce a significant systematical effect, although we note that certain channels (such as $^3P_2$) show marked improvement at NNLO.
While the results for the $D$-waves appear rather accurate, we note that the current
way of smearing the LO contact interactions does produce unwanted additional forces in the $D$-wave channels, which should be dominated by
OPE alone. The $D$-wave channels are expected to improve further upon addition of the
N$^3$LO contributions, which will be included in future work~\cite{Du2017}.

In Table~\ref{fit:procedure}, we also give the value of $\chi^2/N_\mathrm{dof}^{}$ for each of our fits ($a=0.98$~fm),
where $N_\mathrm{dof}^{}$ equals the number of fitted data points (phase shifts or mixing angles at a given momentum) 
minus the number of adjustable parameters. At LO with $a=0.98$~fm, we find $\chi^2/N_\mathrm{dof}^{} \simeq 30$, 
which is reasonable given the rather stringent uncertainty criterion~(\ref{phase_error}) of the PWA. This indicates that 
we have a satisfactory description of the $^1S_0$ and $^3S_1$ channels in the range $p_{\mathrm{CM}}^{} <100$~MeV. 
At NLO, the main contribution to $\chi^2/N_\mathrm{dof}^{}$ arises from $\epsilon_1^{}$ with 
$p_{\mathrm{CM}}^{} > 100$~MeV, while at NNLO $\epsilon_1^{}$ and the $P$-wave channels contribute roughly
equally. These observations are consistent with the results shown in Fig.~\ref{lattice_spacing_98}. 

We also give the $S$-wave low-energy parameters for $a = 0.98$~fm in Table~\ref{Swave_parameter_98}, along with
a summary of the fitted parameters. We find that the NLO and NNLO results clearly provide the closest agreement with 
the empirical scattering lengths and effective ranges, taken from Ref.~\cite{Machleidt:2000ge}. We note that
$a_{^3S_1}^{}$ and $r_{^3S_1}^{}$ are both stable at various orders in NLEFT, and reasonably close to the empirical values. 
This is easily understood since the phase shift in the $^3S_1$ channel is accurately reproduced already at LO. 
For $a_{^1S_0}^{}$ and $r_{^1S_0}^{}$, a clear improvement is observed at NLO and NNLO compared to the results at LO. 
We also find that at NLO and NNLO, $E_d^{}$ can be accommodated 
without sacrificing any accuracy in the other low-energy parameters. Finally, 
$C_{^1S_0}^{}$ and $C_{^3S_1}^{}$ for $a = 0.98$~fm are in reasonably close agreement with the continuum results of
Ref.~\cite{Epelbaum:2014efa} for a cutoff of $R=1.0$~fm, which suggests that lattice artifacts are under control.

\subsection{Variation of the lattice spacing \label{ls_dependence}}

Up to this point, we have mostly elaborated on our results for $a = 0.98$~fm, which is the smallest lattice spacing we have considered.
We shall next comment on our findings when the lattice spacing is varied in the range $1.97 \geq a \geq 0.98$~fm, while the physical
lattice volume is kept constant at $V = (La)^3 \simeq (63~\mathrm{fm})^3$ (see Table~\ref{lattice_spacing} for a summary of lattice parameters).
As we work within the transfer matrix formalism, the temporal lattice spacing $a_t^{}$ should also be varied when $a$ is changed.
Here, we choose $a_t^{}$ such that $a_t^{}/a^2$ is kept constant. This is motivated by the fact that the Hamiltonian scales with 
the lattice spacing as $H \sim 1/a^2$. For a pioneering LO calculation of the effects of varying $a$, see also Ref.~\cite{Klein:2015vna}.

In Table~\ref{LECs:uncertainty}, we summarize the fitted constants of the NN interaction as a function of $a$, along with the
$S$-wave low-energy parameters in Table~\ref{Swave:parameter}. We note that the uncertainties of the fitted constants are
obtained by an analysis of the variance-covariance matrix according to Eq.~(\ref{uncertainty_coefficient}), while those of the 
$S$-wave parameters are obtained using Eq.~(\ref{uncertainty_observable}). Our computed $S$-wave parameters appear 
very stable with respect to lattice spacing variation, which suggests that lattice spacing effects 
are small in the $S$-wave channels. 

Our results for neutron-proton phase shifts and mixing angles for
$a = 1.32$~fm are shown in Fig.~\ref{lattice_spacing_132}, for $a = 1.64$~fm in Fig.~\ref{lattice_spacing_164}, and finally
for $a = 1.97$~fm in Fig.~\ref{lattice_spacing_197}. Together with the results for $a = 0.98$~fm shown in Fig.~\ref{lattice_spacing_98}, 
it is immediately apparent that lattice spacing effects are small for the $S$-waves in the range $0 < p_{\mathrm{CM}}^{} < 200$~MeV, 
which is consistent with the behavior of the $S$-wave parameters. On the other hand, this situation is quite
different for the $P$-waves and $D$-waves. For these higher partial waves, as well as for the
mixing angles $\epsilon_1^{}$ and $\epsilon_2^{}$, the lattice spacing effects remain small only up to $p_{\mathrm{CM}}^{} < 100$~MeV. 
For $p_{\mathrm{CM}}^{} >100$~MeV, the deviations from the Nijmegen PWA increase rapidly, but are nevertheless systematically reduced
when $a$ is decreased.
 
To conclude, for the $S$-waves the lattice spacing effects remain small throughout the range of $p_{\mathrm{CM}}^{}$ considered here,
even for the (rather coarse) lattice spacing of $a =1.97$~fm. For the $P$-waves and $D$-waves, this situation holds only up to 
$p_{\mathrm{CM}}^{} \simeq 100$~MeV. However, we note that $a = 0.98$~fm suffices to give an accurate description 
for $p_{\mathrm{CM}}^{} \simeq 200$~MeV, regardless of the channel under consideration. This suggests that the observed 
discrepancies could be eliminated by a combination of improved lattice momentum operators and N3LO effects, possibly taken together
with a lattice spacing somewhat smaller than $a =1.97$~fm. We would like to stress  that the phase shifts agree within uncertainties
below 150~MeV (with a few exceptions) for the lattice spacings considered.
This validates the statements made in Ref.~\cite{Klein:2015vna} about the
lattice spacing independence of observables in the two-nucleon sector.


\subsection{Perturbative treatment of higher orders \label{perturbative}}


\begin{table}[t]
\begin{center}
\caption{Summary of fit results (in units of $a$) for the perturbative NLO+NNLO analysis at $a=1.97$~fm.
Fitted values of $E_d^{}$ are indicated by a dagger~($\dagger$). Note that the values of
$C_{^1S_0}^{}$, $C_{^3S_1}^{}$ and $b_s^{}$ are fixed by the LO fit.
\label{Tab:Perturbative_LECs}}
\smallskip
\begin{tabular*}{0.475\textwidth}{@{\extracolsep{\fill}}lrrr}
\hline\hline
\noalign{\smallskip}
    & LO & NLO & NNLO  
\smallskip \\
 \hline
  $C_{^1S_0}^{}$         &   $-0.462(8)$    & $-$    &    $-$    \\
  $C_{^3S_1}^{}$         &   $-0.633(6)$    & $-$    &    $-$    \\
  $b_s^{}$            &   $ 0.054(3)$  & $-$    &    $-$    \\
  $\Delta C$          &  $-$              & $-0.2(2)$      &    $-0.0(2)$    \\
  $\Delta C_{I^2}^{}$    &  $-$              & $-0.02(9)$     &    $0.03(9)$       \\
  $C_{q^2}^{}$           &  $-$              & $0.03(5)$      &    $0.12(5)$     \\
  $C_{I^2, q^2}^{}$      &  $-$              & $0.04(2)$      &    $0.03(3)$      \\
  $C_{S^2, q^2}^{}$      &  $-$              & $-0.05(5)$     &    $-0.02(5)$       \\
  $C_{S^2, I^2, q^2}^{}$ &  $-$              & $0.00(2)$      &    $-0.01(2)$       \\
  $C_{(q\cdot S)^2}^{}$  &   $-$             & $0.06(2)$      &    $-0.05(2)$      \\
  $C_{I^2, (q\cdot S)^2}^{}$ &   $-$         & $-0.10(2)$     &    $-0.07(2)$    \\
  $C_{(q \times S)\cdot k}^{I=1}$ &  $-$  & $0.039(5)$     &    $0.038(5)$     \\
  \hline 
  \noalign{\smallskip}
  $E_d^{}$~[MeV]      &  $-2.02(4)$      &  $-2.224(3)^\dagger$ &    $-2.224(3)^\dagger$   
  \smallskip \\
 \hline\hline                          
\end{tabular*}
\end{center}
\end{table}


\begin{table}[t]
\begin{center}
\caption{Summary of fit results (in units of $a$) for the perturbative NLO+NNLO analysis at $a=1.64$~fm.
Notation as in Table~\ref{Tab:Perturbative_LECs}. 
\label{Tab:Perturbative_LECs120}}
\smallskip
\begin{tabular*}{0.475\textwidth}{@{\extracolsep{\fill}}lrrr}
\hline\hline
\noalign{\smallskip}
    & LO & NLO & NNLO  
\smallskip \\
 \hline
  $C_{^1S_0}^{}$               &  $-0.47(1)$ & $-$ & $-$ \\
  $C_{^3S_1}^{}$               &  $-0.64(1)$ & $-$ & $-$ \\
  $b_s^{}$                  &  $0.091(5)$ & $-$ & $-$ \\
  $\Delta C$                &  $-$         & $-0.2(2)$    &  $0.3(3)$    \\
  $\Delta C_{I^2}^{}$          &  $-$         & $-0.00(9)$   &  $0.1(1)$       \\
  $C_{q^2}^{}$                 &  $-$         & $0.04(6)$    &  $0.18(6)$     \\
  $C_{I^2, q^2}^{}$            &  $-$        & $0.08(3)$    &  $0.06(3)$      \\
  $C_{S^2, q^2}^{}$            &  $-$         & $-0.05(5)$   &  $0.00(6)$       \\
  $C_{S^2, I^2, q^2}^{}$       &  $-$         & $-0.01(3)$   &  $-0.01(3)$       \\
  $C_{(q\cdot S)^2}^{}$        &  $-$         & $0.06(3)$    &  $0.08(4)$      \\
  $C_{I^2, (q\cdot S)^2}^{}$   &  $-$         & $-0.07(3)$   &$-0.06(4)$    \\
  $C_{(q \times S)\cdot k}^{I = 1}$ &  $-$ & $0.051(9)$    &  $0.05(1)$     \\
  \hline 
    \noalign{\smallskip}
  $E_d^{}$~[MeV]            &  $-2.13(4)$ & $-2.224(2)^\dagger$&     $-2.224(2)^\dagger$     
  \smallskip \\
 \hline\hline                          
\end{tabular*}
\end{center}
\end{table}
 
 
\begin{table}[t]
\begin{center}
\caption{Summary of fit results (in units of $a$) for the perturbative NLO+NNLO analysis at $a=1.32$~fm.
Notation as in Table~\ref{Tab:Perturbative_LECs}. 
\label{Tab:Perturbative_LECsa150}}
\smallskip
\begin{tabular*}{0.475\textwidth}{@{\extracolsep{\fill}}lrrr}
\hline\hline
\noalign{\smallskip}
    & LO & NLO & NNLO  
\smallskip \\
 \hline
  $C_{^1S_0}^{}$           &   $-0.44(1)$    & $-$   &  $-$    \\
  $C_{^3S_1}^{}$           &   $-0.59(1)$    & $-$    &  $-$  \\
  $b_s^{}$                      &   $0.18(1)$   & $-$    &  $-$  \\
  $\Delta C$                    &  $-$             & $0.0(2)$     &  $0.4(2)$   \\
  $\Delta C_{I^2}$              &  $-$             & $0.05(9)$    &  $0.30(9)$   \\
  $C_{q^2}$                     &  $-$             & $0.04(6)$    &  $0.62(6)$   \\
  $C_{I^2, q^2}$                &  $-$             & $0.19(4)$    &  $0.07(3)$   \\
  $C_{S^2, q^2}$                &  $-$             & $-0.03(5)$   &  $0.07(5)$   \\
  $C_{S^2, I^2, q^2}$           &  $-$             & $-0.01(3)$   &  $-0.12(3)$  \\
  $C_{(q\cdot S)^2}$            &  $-$             & $0.09(4)$    &  $0.02(4)$  \\
  $C_{I^2, (q\cdot S)^2}$       &  $-$             & $-0.05(4)$   &  $0.12(4)$   \\
  $C_{(q \times S)\cdot k}^{I = 1}$ &  $-$         & $0.12(1)$    &  $0.11(1)$   \\
  \hline 
    \noalign{\smallskip}
  $E_d^{}$~[MeV]                &  $-2.14(3)$     & $-2.224(1)^\dagger$&  $-2.224(1)^\dagger$
  \smallskip \\
 \hline\hline                          
\end{tabular*}
\end{center}
\end{table}


\begin{figure*}[t]
\begin{center}
\includegraphics[width=\textwidth]{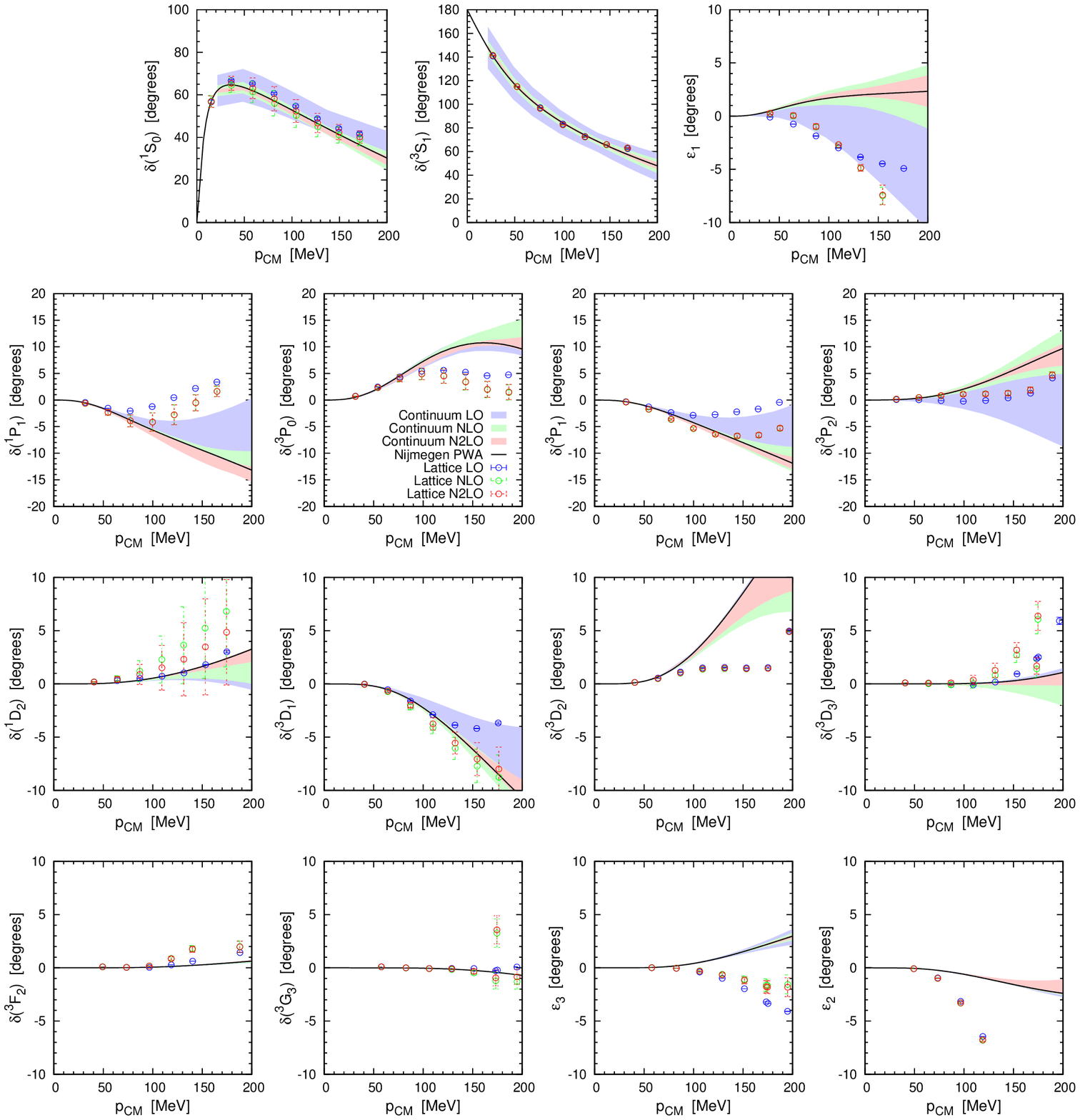}
\caption{Fitted LO + perturbative NLO/NNLO neutron-proton phase shifts and mixing angles for $a = 1.97$~fm. The shaded bands denote the continuum results 
of Ref.~\cite{Epelbaum:2014efa}, and the NPWA is given by the black line.
\label{Fig:Plot_pert}}
\end{center}
\end{figure*}


\begin{figure*}[t]
\begin{center}
\includegraphics[width=\textwidth]{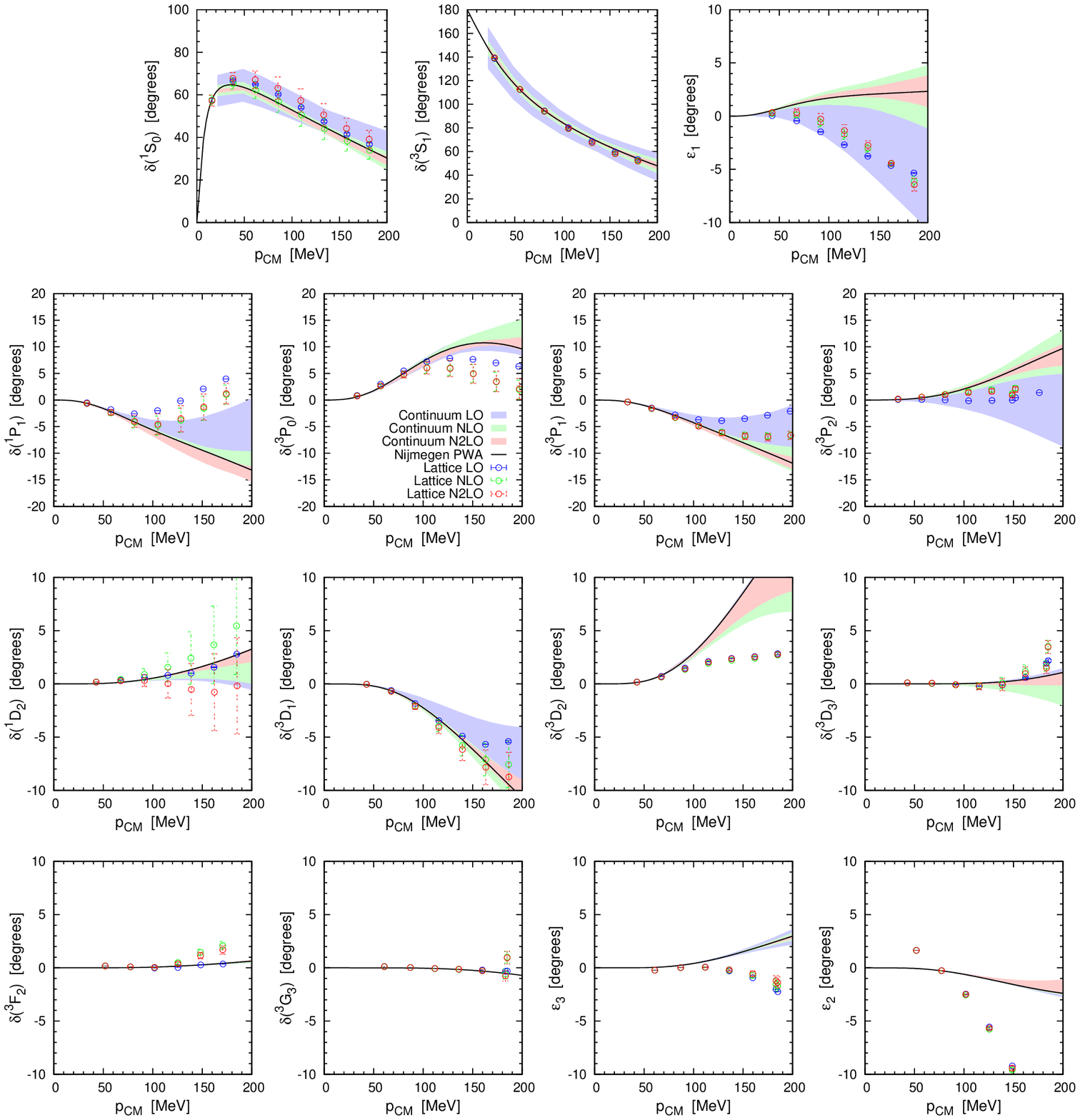}
\caption{Fitted LO + perturbative NLO/NNLO neutron-proton phase shifts and mixing angles for $a = 1.64$~fm. The shaded bands denote the continuum results 
of Ref.~\cite{Epelbaum:2014efa}, and the NPWA is given by the black line. \label{Fig:Plot_pert120}}
\end{center}
\end{figure*}


\begin{figure*}[t]
\begin{center}
\includegraphics[width=\textwidth]{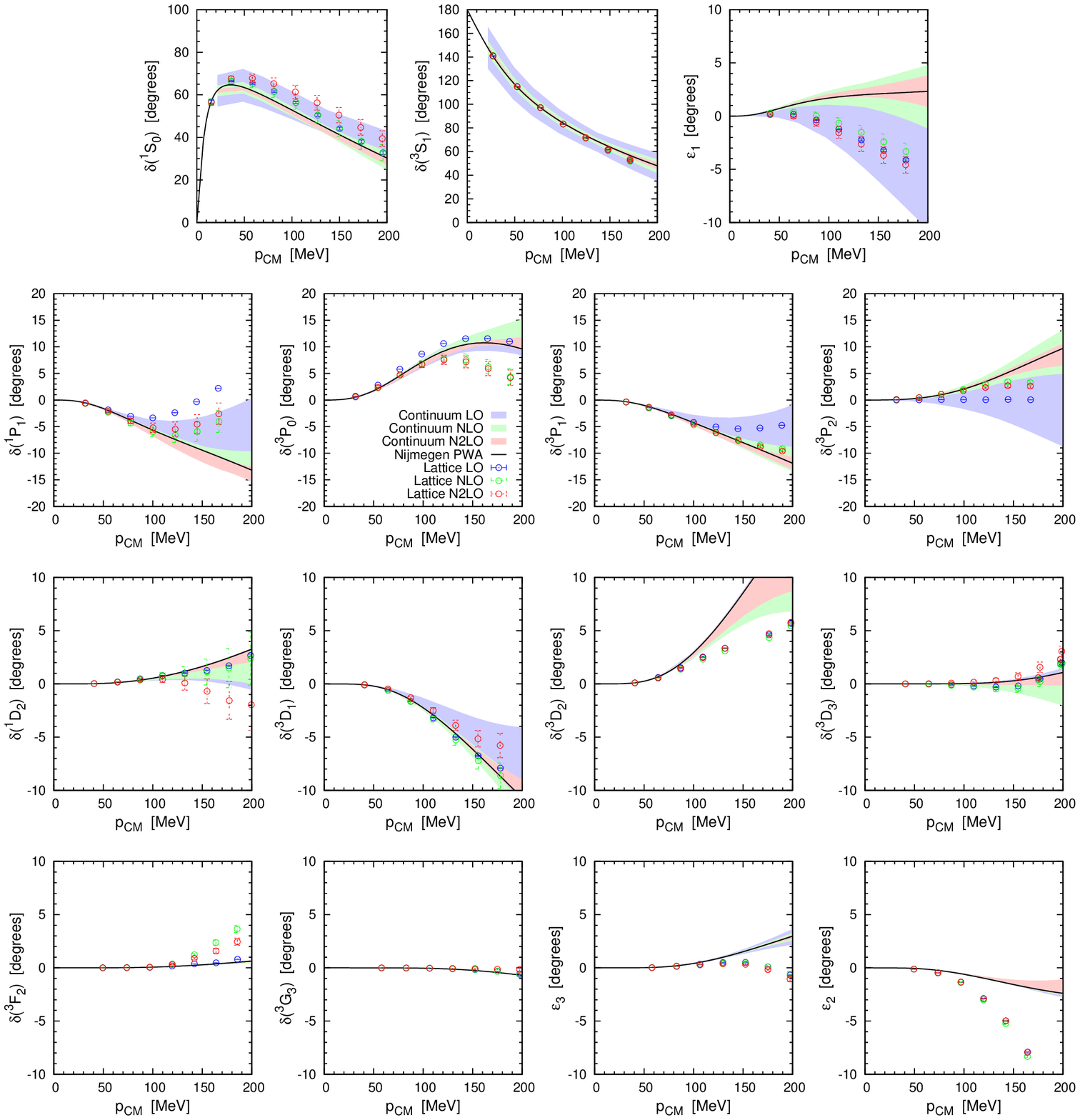}
\caption{Fitted LO + perturbative NLO/NNLO neutron-proton phase shifts and mixing angles for $a = 1.32$~fm. The shaded bands denote the continuum results 
of Ref.~\cite{Epelbaum:2014efa}, and the NPWA is given by the black line. \label{Fig:Plot_pert150}}
\end{center}
\end{figure*}


We have thus far demonstrated that non-perturbative fits to neutron-proton scattering data are feasible to any given order in NLEFT, provided
that the requisite potential operators have been worked out. Nevertheless, for practical reasons (such as sign oscillations and increased 
computational complexity) 
the contributions of NLO and higher orders are usually treated perturbatively in Monte Carlo simulations of nuclear many-body systems. With this in mind, we 
show here how our analysis of phase shifts and mixing angles can be applied in a way consistent with current lattice Monte Carlo work.

Before discussing our results, we briefly summarize the differences between the perturbative and non-perturbative analyses.
We again start with a LO fit, the parameters of which are fixed by fitting the $^1S_0$ and $^3S_1$ channels (but not $E_d^{}$). 
As in the non-perturbative analysis, for the LO fits we consider data up to $p_\mathrm{CM}^\mathrm{max} = 100$~MeV. 
For higher-order (NLO and NNLO) fits, we include data up to $p_\mathrm{CM}^\mathrm{max} = 150$~MeV.
Since higher orders in NLEFT are treated perturbatively, the transfer matrix is constructed in a different way than in 
Eq.~(\ref{Eq:Transfer_matrix_non-perturbative}).
To be specific, in the perturbative analysis the transfer matrix is
\begin{align}
\label{Eq:Transfer_matrix_perturbative}
M^\mathrm{pert} = M_\mathrm{LO}^{} 
- \alpha_t^{} :(V_\mathrm{NLO}^{} + V_\mathrm{NNLO}^{} ) M_\mathrm{LO}^{} :,
\end{align}
where
\begin{align}
M_\mathrm{LO}^{} = \: 
:\exp\big[ -\alpha_t^{} (H_\mathrm{free}^{} + V_\mathrm{LO}^{}) \big] :,
\end{align}
and as in previous Monte Carlo studies of NLEFT, we introduce the additional operators~\cite{Borasoy:2007vi}
\begin{eqnarray}
\Delta V &\equiv& \Delta C  \\
\Delta V_{I^2}^{} &\equiv& \Delta C_{I^2}^{} \, 
\vec{\tau}_1^{} \cdot \vec{\tau}_2^{},
\end{eqnarray}
which we classify as NLO perturbations and add to the NLO potential in Eq.~\eqref{Eq:V_NLO} when $M^{\mathrm{pert}}$ is computed. 
This is done because the LO LECs are kept fixed and thus fitting these finite shifts is  equivalent to a refit of the LO LECs,
as it is done in the non-perturbative case.
Additionally, $\Delta V$ and $\Delta V_{I^2}^{}$ absorb part of the (sizable) short-distance contributions from TPE at NLO and NNLO. 
At NLO, we also studied an operator of the form $\sum_i \tau_{1,i}^{} \tau_{2,i}^{} q_i^2$ which accounts for
rotational symmetry breaking effects on the lattice, but no significant effects were observed.

As for the non-perturbative case, we give results for a range of lattice spacings for the perturbative analysis. The fitted parameters for
$a=1.97$~fm, $a=1.64$~fm and $a=1.32$~fm are given in Tables~\ref{Tab:Perturbative_LECs}, \ref{Tab:Perturbative_LECs120} 
and~\ref{Tab:Perturbative_LECsa150}, respectively. The corresponding phase shifts and mixing angles are shown in 
Figs.~\ref{Fig:Plot_pert}, \ref{Fig:Plot_pert120} and~\ref{Fig:Plot_pert150}.
For each computed phase shift, we provide an estimated uncertainty according to
\begin{equation}
\Delta \delta \equiv 
\sqrt{(J_\delta^T)_i^{} \mathcal{E}_{ij} (J_\delta^{})_j^{}} \times \sqrt{\chi^2_\mathrm{min} / N_\mathrm{dof}^{}},
\label{Delta_delta}
\end{equation}
where $\mathcal{E}_{ij}$ denotes the variance-covariance matrix of the fitted parameters, according to Eq.~(\ref{eq:covmatrix}),
and $J_\delta^{}$ is the Jacobian vector of the phase shift (or mixing angle) in question. The last factor in Eq.~(\ref{Delta_delta})
is the so-called Birge factor described in App.~\ref{app_errors}, which approximately accounts for the systematical errors in the analysis.

At LO, we reproduce well the low-momentum region, and obtain a realistic deuteron binding energy.
In particular, we note that the $^3S_1$ PWA data are almost perfectly reproduced. This is largely caused by the very accurate PWA data of this channel,
which gives this channel a relatively high weight in the $\chi^2$ function. We note that this may potentially worsen the agreement in other channels, where
a comparable accuracy of the PWA data is not available. Also, the expectation is that the $P$-waves should be well described at LO, since they are 
dominated by the OPEP contribution. The reason why this is not the case for our LO results is that, 
in the perturbative calculation, and in order to be consistent with the Monte Carlo simulations, 
we treat the momentum $\vec{q}\,{}^2$ in the denominator of the OPE as in Eq.~(\ref{Eq:q2_OPE}), and factors of 
$\vec{q}$ as in Eq.~(\ref{momentum:a}). This choice considerably suppresses the OPEP contribution already at intermediate momenta, 
which worsens the description of the $P$-waves. 


Moving to NLO, a significant improvement is found in some channels, particularly for $^1P_1$ and $^3P_1$, where the PWA is now well described up 
to $\sim 100$~MeV. On the other hand, 
we note that the $^3P_0$ and $^3P_2$ channels, as well as the $D$-waves, show little improvement. We attribute these features to the deficiencies 
in the OPE as mentioned above.
The description of the $S$-wave channels is found to improve at intermediate momenta, which is mainly due to the NLO contact terms and to the 
parts of the NLO TPEP that contribute to the $S$-waves.

At NNLO, while no new unknown parameters contribute, the sub-leading TPEP enters as a prediction from $\pi N$ scattering in Chiral EFT. 
Thus, the NLO constants are refitted at NNLO in order to absorb the strong short-distance isoscalar contributions from the $\pi N$ LECs. 
The NLO and NNLO results appear in most cases virtually indistinguishable (as shown in Fig.~\ref{Fig:Plot_pert}) as far as the level of agreement
with the PWA is concerned, except for the $^1D_2$ channel where the high-momentum tail is noticeably improved.

For our perturbative analysis, we have also compared the computed scattering observables at different orders in NLEFT with the continuum results 
of Ref.~\cite{Epelbaum:2014efa}.
We find that our $S$-waves agree with the continuum results (within errors) up to at least $p_\mathrm{CM}^{} \simeq 100$~MeV, 
and in some cases over the entire range of momenta considered. The $P$-waves show good agreement within errors only for some channels, and 
only for NLO/NNLO.  As already mentioned, this is mainly due to the non-optimal description of OPE at LO. For the $D$-wave channels, 
only $^3D_1$ shows good agreement  with the continuum calculations. For the $^1D_2$ channel, the LO and NLO results overshoot the 
continuum error band, while the NNLO result is in agreement due to the large uncertainty. 
For $^3D_2$, the NLO/NNLO terms do not contribute at all and hence cannot improve the result. 
Further, for $^3D_3$ the lattice calculations start to deviate from the PWA and the continuum results for $p_\mathrm{CM}^{} > 100$~MeV.

Finally, it is important to stress that for cms momenta below 150~MeV, the phase shifts agree within the uncertainties (with the exception
of $\epsilon_1$, were deviations set in at about 110~MeV). This validates the statements made in Ref.~\cite{Klein:2015vna} about the
lattice spacing independence of observables in the two-nucleon sector.

\subsection{Further improvements} 

 
\begin{table}[t]
\begin{center}
\caption{Summary of fit results with perturbatively 
improved OPE (in units of $a$) for the perturbative NLO+NNLO analysis at $a=1.97$~fm.
Notation as in Table~\ref{Tab:Perturbative_LECs}. 
\label{Tab:Perturbative_LECsDe_DX}}
\smallskip
\begin{tabular*}{0.475\textwidth}{@{\extracolsep{\fill}}lrrr}
\hline\hline
\noalign{\smallskip}
    & LO & NLO & NNLO  
\smallskip \\
 \hline
  $C_{^1S_0}^{}$         &   $-0.462(8)$    & $-$    &    $-$    \\
  $C_{^3S_1}^{}$         &   $-0.633(6)$    & $-$    &    $-$    \\
  $b_s^{}$            &   $ 0.054(3)$  & $-$    &    $-$    \\
  $\Delta C$          &  $-$             & $-0.2(3)$      &    $-0.0(3)$    \\
  $\Delta C_{I^2}^{}$    &  $-$              & $-0.1(1)$      &    $0.03(9)$       \\
  $C_{q^2}^{}$           &  $-$              & $-0.03(7)$     &    $0.05(7)$     \\
  $C_{I^2, q^2}^{}$      &  $-$             & $0.09(3)$      &    $0.06(3)$      \\
  $C_{S^2, q^2}^{}$      &  $-$              & $-0.05(6)$     &    $0.00(6)$       \\
  $C_{S^2, I^2, q^2}^{}$ &  $-$              & $0.00(2)$      &    $-0.03(3)$       \\
  $C_{(q\cdot S)^2}^{}$  &   $-$             & $0.02(2)$      &    $-0.03(3)$      \\
  $C_{I^2, (q\cdot S)^2}^{}$ &   $-$         & $-0.07(2)$     &    $0.10(3)$    \\
  $C_{(q \times S)\cdot k}^{I=1}$ &  $-$  & $0.014(7)$     &    $0.012(5)$     \\
  \hline 
    \noalign{\smallskip}
  $E_d^{}$~[MeV]      &  $-2.02(4)$      &  $-2.224(3)^\dagger$ &    $-2.224(3)^\dagger$ 
  \smallskip \\
 \hline\hline                          
\end{tabular*}
\end{center}
\end{table}


\begin{figure*}[t]
\begin{center}
\includegraphics[width=\textwidth]{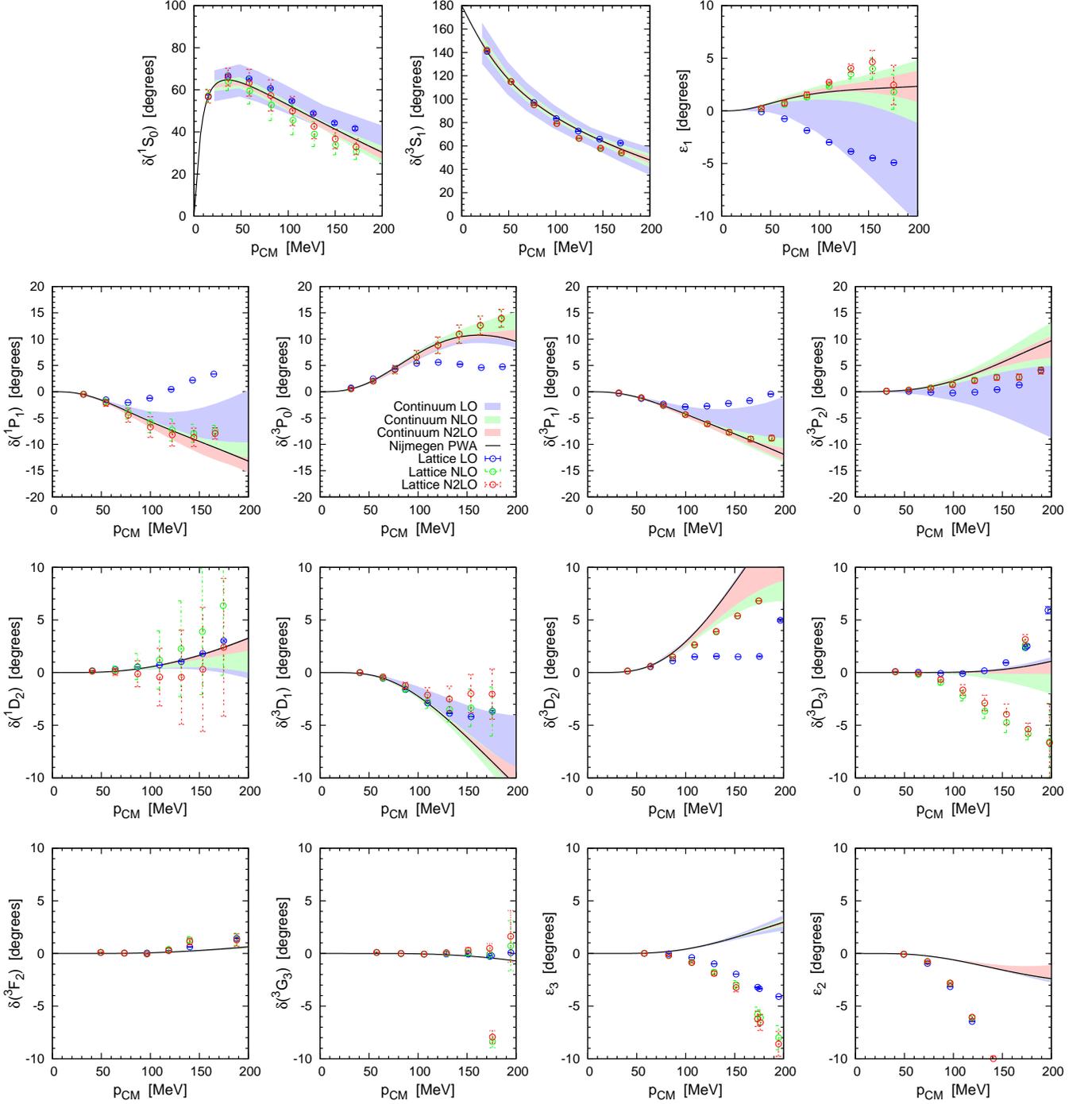}
\caption{Fitted LO + perturbative NLO/NNLO neutron-proton phase shifts and mixing angles for $a = 1.97$~fm including the improved OPE. The shaded bands denote the continuum results 
of Ref.~\cite{Epelbaum:2014efa}, and the NPWA is given by the black line.
\label{Fig:Plot_pert_De_DX}}
\end{center}
\end{figure*}


Next, we shall discuss two problems that require further study to resolve. First, 
while a clear improvement was observed in the non-perturbative case for the scattering observables (and fitted parameters) as $a$ 
was decreased, a similar improvement is not found in the
perturbative analysis. At LO, the quality of the description improves in general with decreasing $a$, particularly for the $^3S_1$-$^3D_1$ 
coupled channel. However, at NLO/NNLO the picture is 
more complicated. We note that the $P$-waves and remaining $D$-wave channels do improve, but the description of the $^1S_0$ and $^3S_1$-$^3D_1$ 
channels may in fact deteriorate for smaller $a$. We attribute this effect to the increasing influence of the TPE potential. While the effect of TPE 
on the $S$-waves can be absorbed by smeared contact 
interactions as was done in the non-perturbative calculation, in the perturbative case we only have standard (without smearing) contact 
interactions available.  This is sufficient for $a=1.97$~fm, as the TPEP contribution then closely resembles a contact interaction. A possible 
solution for smaller $a$ would be to include a smeared version of 
the NLO/NNLO contact interactions. Alternatively, one could use exact momentum operators for the NLO contact terms, which do have 
a higher influence at larger momenta. 
This was not necessary nor observable for $a=1.97$~fm, but may improve the $^3S_1$ channel once $a$ is decreased. Finally, 
we note that the choice of $c_1^{}$, $c_3{}$ and $c_4^{}$ may also have an effect, as it influences the strength of the different contribution to TPEP.
However, to use the full power of chiral EFT, one should utilize the values determined from pion-nucleon scattering.

Second, we show preliminary results including a perturbative improvement of the OPE operator. In order to remedy the aforementioned 
discrepancies in the peripheral partial
waves such that consistency with the Monte Carlo calculation is maintained, we introduce a new operator 
at NLO that accounts for the difference between OPEP with the momenta of Eqs.~(\ref{Eq:q2_OPE}) and~(\ref{momentum:a})
and the ``exact'' lattice momentum $\vec{q}_\mathrm{ex}^{} \equiv 2 \pi \vec{k} / L$.
This gives
\begin{eqnarray}
\Delta V_\mathrm{OPE}^{} &\equiv& 
-\frac{g_A^2}{4F_\pi^2} \, {\bf \tau}_1^{} \cdot {\bf \tau}_2^{} \\
&\times&
\Bigg[ \frac{(\vec{\sigma}_1^{} \cdot \vec{q}_\mathrm{ex}^{}) (\vec{\sigma}_2^{}\cdot \vec{q}_\mathrm{ex}^{})}
{\vec{q}\,^2_\mathrm{ex} + M_\pi^2}
- \frac{(\vec{\sigma}_1^{}\cdot \vec{q}\,) (\vec{\sigma}_2^{}\cdot \vec{q}\,)}{\vec{q}\,{}^2 + M_\pi^2} \Bigg],
\nonumber
\end{eqnarray} 
so that by adding $\Delta V_\mathrm{OPE}^{}$ to $V_\mathrm{OPE}^{}$, one recovers OPEP with the exact momentum. It should be noted that 
this differs slightly from treating
OPEP at LO with the exact momentum, since $\Delta V_\mathrm{OPE}^{}$ is treated as a perturbation, while $V_\mathrm{OPE}^{}$ is implemented 
non-perturbatively. 
Also, $\vec{q}$ approaches $\vec{q}_\mathrm{ex}^{}$ as $a \to 0$. This means that, simultaneously, $\Delta V_\mathrm{OPE}^{}$ becomes less important, 
and $V_\mathrm{OPE}^{}$ gives a better description of the $P$-waves, as we approach the continuum limit. This is consistent with 
Figs.~\ref{lattice_spacing_132}-\ref{lattice_spacing_197} of the non-perturbative calculation, where the $P$-waves clearly improve as $a$ decreases.

Our perturbative results with $\Delta V_\mathrm{OPE}^{}$ included are given in Fig.~\ref{Fig:Plot_pert_De_DX}
and Table~\ref{Tab:Perturbative_LECsDe_DX}, where as expected 
one can observe a clear improvement in the description of the $P$-waves. The experimental results for the $^1P_1$, $^3P_0$ and $^3P_1$ channels are now well
reproduced for the range of fitted momenta $p_\mathrm{CM}^{} < 150$~MeV. In general, we find that 
all the $P$-wave channels and the $\epsilon_1^{}$ mixing angle appear much closer to the PWA at NLO with improved OPE, than without this correction. 
Additionally, we find that the $D$-waves (except for the $^3D_3$ channel) also improve significantly with respect to the LO result. 
In the case of $^3D_3$, the correction is too large and so the computed values fall below the PWA ones. 
Again, this improvement is mostly attributable to $\Delta V_\mathrm{OPE}^{}$, although we recall that the leading (NLO) TPEP 
also contributes to the high-momentum tails in some of the $D$-wave channels. 
 
\subsection{Nuclear binding energies \label{propagation}}

In Monte Carlo simulations of NLEFT, the binding energies of nuclei receive perturbative energy shifts that depend on the
NLO constants $C_i^{}$ and their uncertainties, in addition to any inherent Monte Carlo uncertainties. For instance, in
Ref.~\cite{Lahde:2013uqa}, only the Monte Carlo errors were taken into account, and the $C_i^{}$ were assumed to be
accurately known and uncorrelated. Since our analysis  provides us with the complete variance-covariance
matrix of the NLO parameters $C_i^{}$, we are now in a position to estimate the uncertainties of the nuclear
binding energies at NNLO, due to uncertainties and correlations of the $C_i^{}$. 
From our present results, we observe larger correlations between $\Delta C$
and $C_q^2$, between $\Delta C_I^2$ and $C_{I^2,q^2}$, between $\Delta C_I^2$ and
$C_{S^2,q^2}$, and also between $C_{(q\cdot S)^2}$ and $C_{I^2,(q\cdot S)^2}$.

In order to obtain a first, rough estimate of the relative magnitude of Monte Carlo and
fitting errors in calculations
of nuclear binding energies $E_B^{}$, we recall that these are calculated according to
\begin{equation}
\label{Eb_NNLO}
E_B^\mathrm{NNLO} = E_B^\mathrm{LO} + C_i^{} \frac{\partial E_B^{}}{\partial C_i^{}} \bigg
|_{C_i^{} = 0},
\end{equation}
where summation over $i$ is assumed. In the Monte Carlo calculation, the LO binding energies
are computed 
non-perturbatively, and the second term in Eq.~(\ref{Eb_NNLO}) represents the perturbative
shift due to the NLO
constants $C_i^{}$ in the 2NF, which we take from Ref.~\cite{Lahde:2013uqa}. We note that
\begin{equation}
E_B^\mathrm{NNLO} \equiv E_B^{}(C_{^1S_0}^{},C_{^3S_1}^{}, b_s^{}, C_i^{}),
\end{equation}
is a function of all the coupling constants up to NNLO, while the LO values
\begin{equation}
E_B^\mathrm{LO} \equiv E_B^{}(C_{^1S_0}^{},C_{^3S_1}^{}, b_s^{}, C_i^{} = 0),
\end{equation}
equal the binding energies at $C_i^{} = 0$. In terms of the 
variance-covariance matrix from the perturbative analysis in
Section~\ref{perturbative}, 
\begin{equation}
\label{delta_B}
\Delta E_B^\mathrm{NNLO} = \sqrt{
\frac{\partial E_B^{}}{\partial C_i^{}} \bigg |_{C_i^{} = 0}
\mathcal{E}_{ij}^{} \: \frac{\partial E_B^{}}{\partial C_j^{}} \bigg |_{C_j^{} = 0}},
\end{equation}
gives us the uncertainties in the NNLO energy shifts due to the fitting errors of the
$C_i^{}$.
The results so obtained are given in Table~\ref{Tab:ErrorProp}.

%
%

We note that the errors due to the uncertainties in the $C_i^{}$ are of comparable magnitude to the Monte Carlo errors,
even when $\mathcal{E}_{ij}^{}$ has been evaluated without consideration of the systematical errors encoded by the 
Birge factor. This may suggest that the procedure of fixing the $C_i^{}$ from two-nucleon data may, at present, be the
main factor limiting the accuracy of NLEFT calculations beyond LO for heavier nuclei. This issue is currently under
further investigation. It should also be noted that the quoted NLEFT binding energies in Table~\ref{Tab:ErrorProp} are
not expected to coincide with the empirical ones, as the $3N$ and higher-order contributions have been neglected (see
Ref.~\cite{Lahde:2013uqa} for further discussion).




\begin{table}[htb]
\begin{center}
\caption{Nuclear binding energies with 2N forces up to NNLO in the NLEFT expansion for $a =
1.97$~fm,
data taken from Ref.~\cite{Lahde:2013uqa}. The first parenthesis gives the
estimated Monte Carlo error in the calculation of $E_B^\mathrm{NNLO}$, and the
second parenthesis the error due to variance-covariance matrix in Eq.~(\ref{delta_B}). For
reference,
we also show the experimental binding energies.
\label{Tab:ErrorProp}}
\smallskip
\begin{tabular}{@{\extracolsep{\fill}}lcc}
\hline\hline
\noalign{\smallskip}
&   $E_B^\mathrm{NNLO}$(2N)          &  $E_B^{}$(exp) 
\smallskip \\
\hline
 $^4$He                         &   $-25.60(6)(2)$   & $-28.30$  \\
 $^8$Be                         &   $-48.6(1)(3)$    & $-56.35$  \\
 $^{12}$C                       &   $-78.7(2)(5)$    & $-92.16$  \\
 $^{16}$O                       &   $-121.4(5)(7)$   & $-127.62$ \\
 $^{20}$Ne                      &   $-163.6(9)(9)$   & $-160.64$ \\
 $^{24}$Mg                      &   $-208(2)(2)$     & $-198.26$ \\
 $^{28}$Si                      &   $-275(3)(2)$     & $-236.54$ \\
\hline\hline
\end{tabular}
\end{center}
\end{table}


\section{Summary \label{summary}}

We have revisited the problem of neutron-proton scattering in NLEFT using the recently developed radial Hamiltonian method. 
For the first time, this has allowed us to perform a 
comprehensive and systematical analysis of neutron-proton phase shifts and mixing angles up to NNLO in the EFT expansion, and at several 
different lattice spacings in the range $1-2$~fm. 
We have also presented a comparison of fully non-perturbative NNLO calculations with a perturbative treatment of contributions beyond LO. 
Decreasing the lattice spacing to $a \sim 1$~fm
necessitated the inclusion of TPEP at NLO and NNLO, and as a consequence the latter is now distinct from the NLO treatment, 
although no new adjustable two-nucleon parameters are introduced at NNLO.

By decreasing the lattice spacing $a$, we have found that a much improved description of neutron-proton scattering can be obtained for 
larger center-of-mass momenta. By considering
the lattice spacings $a=1.97$~fm, $a = 1.64$~fm, $a = 1.32$~fm and $a = 0.98$~fm, we found that $a=1.97$~fm provides a good description up to 
$p_\mathrm{CM}^{} \simeq 100$~MeV, whereas the results for 
$a = 0.98$~fm are reliable up to $p_\mathrm{CM}^{} \simeq 200$~MeV. In general, the systematical errors are much reduced as smaller lattice spacings.
Our results suggest that the range of applicability in most channels could be significantly extended by improving the lattice 
momenta in the OPE and NLO contact interactions. Most importantly, however, is the finding that for momenta $p_\mathrm{CM}^{} \lesssim 100$~MeV,
the physics of the two-nucleon system is independent of the lattice spacing $a$, when $a$ is varied in the range from 1~fm to 2~fm. Furthermore,
we have also investigated the error propagation of the uncertainties of the four-nucleon LECs into the binding energies of 
alpha-type nuclei up to $^{28}$Si.

There are several directions in which the present work should be extended. The inclusion of N3LO contact terms and TPE contributions is 
underway and will appear in  a separate publication, along with the inclusion of electromagnetic effects. Also, the preliminary error analysis 
presented here will be investigated further in a subsequent publication,
especially for the propagation of the variances and covariances of the fitted NLO constants to the nuclear binding energies.
Also, as pointed out in Ref.~\cite{Elhatisari:2016owd}, different smearing procedures and fitting also to scattering processes with
more nucleons allows one to taylor interactions that might be preferable in larger nuclear systems.


\begin{acknowledgments}

We are grateful to Serdar Elhatisari, Evgeny Epelbaum and Hermann Krebs for useful discussions.
We acknowledge partial financial support from the Deutsche Forschungsgemeinschaft (Sino-German CRC 110),
BMBF (Grant No. 05P12PDTEE), the U.S. Department of Energy, Office of Science, Office of Nuclear Physics under contracts 
DE-FG02-03ER41260 and DE-AC05-06OR23177, the Magnus Ehrnrooth Foundation of the Finnish Society of
Sciences and Letters, MINECO (Spain), and the ERDF (European Commission) grant FPA2013-40483.
The work of UGM was also supported by the Chinese Academy of Sciences (CAS) President's International 
Fellowship Initiative (PIFI) (Grant No. 2017VMA0025).
\end{acknowledgments}


\appendix

\section{Density and current operators \label{app_operators}}

Here, we define the various nucleon density and current operators that we use throughout our discussion of the nucleon-nucleon
interaction in the main text. Following Refs.~\cite{Borasoy:2006qn,Borasoy:2007vi}, we define the local density operator
\begin{equation}
\rho(\vec{n}) \equiv \sum_{i,j} a_{i, j}^\dagger(\vec n) a_{i, j}^{}(\vec n), 
\end{equation}
the local isospin density operator
\begin{equation}
\rho_I^{}(\vec n) \equiv \sum_{i, j, j^\prime} a_{i, j}^\dag(\vec{n}) (\tau_I)_{j, j^\prime} a_{i, j^\prime}(\vec{n}), 
\end{equation} 
the local spin density operator
\begin{equation}
\rho_S^{}(\vec n) \equiv \sum_{i, i^\prime, j} a_{i, j^\prime}^\dag(\vec{n}) (\sigma_S)_{i, i^\prime} a_{i^\prime, j} (\vec{n}), 
\end{equation}
and the local isospin-spin density operator
\begin{equation}
\rho_{S,I}^{}(\vec n) \equiv \sum_{i, i^\prime, j, j^\prime} a_{i, j}^\dag(\vec{n}) (\sigma_S)_{i, i^\prime} (\tau_I)_{j, j^\prime} a_{i^\prime, j^\prime}(\vec{n}),
\end{equation}
where $\sigma_S^{}$ and $\tau_I^{}$ denote the Pauli matrices for spin and isospin, respectively.
Similarly, we define the current density operator
\begin{equation}
\Pi_l^{}(\vec n) \equiv \sum_{i,j} a_{i,j}^\dagger(\vec n) \nabla_l^{} a_{i,j}^{}(\vec n) 
- \sum_{i,j} \nabla_l^{} a_{i,j}^\dagger(\vec n) a_{i,j}^{}(\vec n),
\end{equation}
the isospin-current density operator
\begin{eqnarray}
\Pi_{l, I}^{}(\vec n) &\equiv& \sum_{i,j,j^\prime} 
a_{i,j}^\dagger(\vec n)(\tau_I^{})_{j,j^\prime}^{} \nabla_l^{} a_{i,j^\prime}^{}(\vec n) \nn \\ 
&-& \sum_{i,j,j^\prime} \nabla_l^{} a_{i,j}^\dagger(\vec n)(\tau_I^{})_{j,j^\prime}^{} a_{i,j^\prime}^{}(\vec n), 
\end{eqnarray}
the spin-current density operator
\begin{eqnarray}
\Pi_{l,S}^{}(\vec n) &\equiv& \sum_{i,i^\prime,j}
a_{i,j}^\dagger(\vec n)(\sigma_S^{})_{i,i^\prime}^{} \nabla_l^{} a_{i^\prime,j}^{}(\vec n) \nn \\
&-& \sum_{i,i^\prime,j} \nabla_l^{} a_{i,j}^\dagger(\vec n)(\sigma_S^{})_{i,i^\prime}^{} a_{i^\prime, j}^{}(\vec n),
\end{eqnarray}
and the spin-isospin-current density operator
\begin{eqnarray}
\Pi_{l,S,I}^{}(\vec n) &\equiv& \sum_{i,i^\prime,j,j^\prime} 
a_{i,j}^\dagger(\vec n) (\sigma_S^{})_{i,i^\prime}^{} (\tau_I^{})_{j,j^\prime}^{} \nabla_l^{} a_{i^\prime, j^\prime}^{}(\vec n) \nn \\
&-& \sum_{i,i^\prime,j,j^\prime} \nabla_l^{} a_{i,j}^\dagger(\vec n) (\sigma_S^{})_{i,i^\prime}^{} (\tau_I^{})_{j,j^\prime}^{} a_{i^\prime,j^\prime}^{}(\vec n),
\qquad
\end{eqnarray}
and we recall that the current operators are used in the isospin-projected spin-orbit term~(\ref{NLO_7}), with $\nabla_l^{}$ defined according to
Eq.~(\ref{definition_nabla}).


\section{Uncertainty analysis \label{app_errors}}

From the definition of $\chi^2$ given in Eq.~(\ref{chisquare}), we note that $\chi^2$ is a 
function of the LO and NLO coupling constants
\begin{equation}
\chi^2 \equiv \chi^2(C_{^1S_0}^{}, C_{^3S_1}^{}, \widetilde C_1^{}, \ldots, \widetilde C_7^{}),
\end{equation}
such that if $\chi^2$ is expanded around its minimum, one finds 
\begin{equation}
\chi^2 = \chi^2_\mathrm{min} + \frac{1}{2} \sum_{i, j} h_{ij}^{} 
(C_i^{} - C_i^\mathrm{min})(C_j^{} - C_j^\mathrm{min}) + \ldots,
\end{equation}
where the Hessian matrix is given by
\begin{equation}
h_{ij}^{} \equiv \frac{\partial^2 \chi^2}{\partial C_i^{} \partial C_j^{}},
\end{equation}
and $C_i^\mathrm{min}$ denotes the set of parameters that minimizes the $\chi^2$ function.
Given that $\chi^2$ reaches its minimum value for $C_i^{} = C_i^\mathrm{min}$, the terms with one derivative vanish. 
Keeping terms up to second order, we obtain the Hessian approximation to the error (or variance-covariance) matrix
\begin{equation}
\mathcal{E}_{ij}^{} \equiv \frac{1}{2} h_{ij}^{-1},
\label{eq:covmatrix}
\end{equation}
and the standard deviations 
\begin{equation}
\sigma_i^{} = \sqrt{\sigma_i^2} = \sqrt{\mathcal{E}_{ii}^{}},
\label{uncertainty_coefficient}
\end{equation} 
of the fitted constants
are obtained from the diagonal elements of the error matrix.

In the absence of systematical errors, we expect to find a normalized chi-square of 
$\tilde\chi^2 \equiv \chi^2/N_\mathrm{dof}^{} \approx 1$, where $N_\mathrm{dof}^{}$ is the number of degrees of freedom
(number of fitted data - number of free parameters) in the fit. However, in our analysis $\tilde\chi^2 > 1$ in most cases, particularly
at LO and for larger values of the lattice spacing $a$. Such a systematical error suggests that the uncertainties computed from
Eq.~(\ref{uncertainty_coefficient}) are underestimated. Following Ref.~\cite{Perez:2014yla}, we therefore rescale the input errors
by the Birge factor~\cite{birge}, according to
\begin{equation}
\Delta_i^{} \rightarrow \Delta_i^{} \sqrt{\tilde\chi^2_\mathrm{min}},
\end{equation}
which leads to the replacement
\begin{equation}
\chi^2 \rightarrow \frac{\chi^2}{\tilde \chi^2_\mathrm{min}} = N_\mathrm{dof}^{} \frac{\chi^2}{\chi^2_\mathrm{min}},
\end{equation}
such that $\chi^2/N_\mathrm{dof}^{} \approx 1$ for $C_i^{} = C_i^\mathrm{min}$.
For a given observable $\mathcal{O}$, we assign an uncertainty according to
\begin{equation}
\Delta \mathcal{O} \equiv \sqrt{(J_\mathcal{O}^{T})_i^{} \mathcal{E}_{ij}^{} (J_\mathcal{O}^{})_j^{}}, 
\label{uncertainty_observable} 
\end{equation}
where 
\begin{equation}
(J_{\mathcal O})_i^{} \equiv \frac{\partial \mathcal{O}}{\partial C_i^{}},
\end{equation}
is the Jacobian vector of $\mathcal{O}$ with respect to the $C_i^{}$.



\end{document}